\documentclass[twocolumn]{bmcart}

\usepackage{amsthm,amsmath,amssymb}
\usepackage[utf8]{inputenc} 

\usepackage{graphicx,subfigure}
\usepackage{hyperref}
\usepackage{xspace}
\usepackage{cleveref}
\usepackage[english]{babel}
\usepackage{booktabs}
\usepackage{datetime}

\startlocaldefs
\newcommand{\refeq}[1]{\cref{eq:#1}}
\newcommand{\hosp}{\textsl{hosp}\xspace}
\newcommand{\conf}{\textsl{conf}\xspace}
\newcommand{\malawi}{\textsl{malawi}\xspace}
\newcommand{\baboons}{\textsl{baboons}\xspace}
\newcommand{\about}{\ensuremath{\simeq}}
\newcommand\mean[1]{\ensuremath{\langle#1\rangle}}
\newcommand\Ncon{\ensuremath{N_{int}}}

\newcommand{\pte}{p.t.e\xspace}
\newcommand\sect[1]{Sect. \ref{sec:#1}}
\newcommand\Fig[1]{Figure \ref{fig:#1}}
\newcommand\Table[1]{Table \ref{tab:#1}}
\newcommand{\SI}{\textit{SI}}
\newcommand{\Sup}[1]{\SI \textit{Appendix, #1}\xspace}
\newcommand{\safeurl}[1]{\href{#1}{#1}}

\newcommand{\corr}[1]{{#1}}

\endlocaldefs

\begin{document}

\begin{frontmatter}

\begin{fmbox}
\dochead{Research}


\title{On the duration of face-to-face contacts}


\author[
   addressref={aff1,aff2},                   
   corref={aff1,aff2},                       
   email={stephane.plaszczynski@ijclab.in2p3.fr}   
]{\inits{SP}\fnm{St\'ephane} \snm{Plaszczynski}}
\author[
     addressref={aff3},                   
     email={gilberto.nakamura@gmail.com}
]{\inits{GN}\fnm{Gilberto} \snm{Nakamura}}
\author[
   addressref={aff1,aff2},                   
   email={basigram@gmail.com}
]{\inits{BG}\fnm{Basile} \snm{Grammaticos}}
\author[
   addressref={aff1,aff2},                   
   email={mathilde.badoual@ijclab.in2p3.fr}
]{\inits{MB}\fnm{Mathilde} \snm{Badoual}}


\address[id=aff1]{
  \orgname{Universit\'e Paris-Saclay, CNRS/IN2P3, IJCLab}, 
  \postcode{91405},                                
  \city{Orsay},                              
  \cny{France}                                    
}
\address[id=aff2]{%
  \orgname{Universit\'e Paris-Cit\'e, IJCLab},
  \postcode{91405},                                
  \city{Orsay},                              
  \cny{France}   
}
\address[id=aff3]{%
  \orgname{Center for Interdisciplinary Research in Biology (CIRB), Coll\`ege de France, CNRS, INSERM, Universit\'e PSL},
  \city{Paris},                              
  \cny{France}   
}



\end{fmbox}

\begin{abstractbox}

\begin{abstract} 
\corr{The analysis of social networks, in particular those describing
face-to-face interactions between individuals, is complex due to
the intertwining of the topological and temporal aspects.
We revisit here both, using public data recorded by the
\textit{sociopatterns} wearable sensors in some very different
sociological environments, putting particular emphasis on the contact duration timelines.}
As well known, the distribution of the contact duration for
all the interactions within a group is broad, 
with tails that resemble each other, but not precisely, in different contexts. 
By separating each interacting pair, we find that the 
\textit{fluctuations} of the contact duration around the
mean-interaction time follow however a very similar pattern.
This common robust behavior is observed on 7 different datasets.
It suggests that, although the set of persons we interact with and the 
mean-time spent together, depend strongly on the environment, 
our tendency to allocate more or less time than usual with a given
individual is invariant, i.e. governed by some rules 
that lie outside the social context.
Additional data reveal the same fluctuations in a baboon population.
This new metric, which we call the relation ``contrast'', can be used to build and test 
agent-based models, or as an input for describing long duration
contacts in epidemiological studies. 
\end{abstract}




\end{abstractbox}
%

\end{frontmatter}


\section*{Introduction}

Since the advent of the Internet, the quantity of digital data describing our behavior
has inflated, offering to scientists an unprecedented
opportunity to study human interactions in a more 
quantitative way. This 
 opened the field of sociology to data-analysis and 
from the hard-science community, came the tacit idea that 
several aspects of the complex human behavior can be modeled  
\cite{Song:2010,Stehle:2010,Zhao:2011,Starnini:2013,Sekara:2016,Flores:2018}. 
With the rapid development of mobile technologies (GPS, Bluetooth,
cellphones) a lot of effort was first put in trying to
capture the patterns of human mobility (for a review, see \cite{Barbosa:2018}).
A more local picture of our everyday social interactions can be obtained
using dedicated proximity sensors.
Following a pioneering experiment that equipped conference participants
with pocket switched devices \cite{Hui:2005,Scherrer:2008}, 
the \textit{sociopatterns} collaboration
(\safeurl{www.sociopatterns.org}) developed some wearable sensors
that allow to register the complex patterns of face-to-face
interactions \cite{Cattuto:2010,Barrat:2014}. The radio-frequency
signal is only recorded if two individual
are in front of each other for a duration of a least 20 s (which is
the timing resolution). We note that, from a sociological point of view,
a distance below 1.5 m covers the traditional 
\textit{private} ($<50$ cm), \textit{personal} ($<$1.2 m) and
\textit{social} ($<3.5$ m) zones. 
The goal is not only to analyze social interactions but also to
understand how information (or a disease) spreads over a real dynamical
network \cite{Stehle:2011,Isella:2010,Starnini:2012,Genois:2015}.
Those sensors were worn by volunteers in several work-related environments: 
scientific conferences \cite{Cattuto:2010,Isella:2010,Stehle:2011},
a hospital ward \cite{Vanhems:2013}, an office \cite{Genois:2015} and at
school \cite{Fournet:2014,Mastrandrea:2015}. 
As part of a UNICEF program, they were also used to
characterize social exchanges in small villages in Kenya and Malawi \cite{Kiti:2016,Ozella:2021}
and for ethological studies on baboons \cite{Gelardi:2020}.

It has been known for a long time that the \textit{overall} distribution of the duration of
contacts in face to face interactions is ``broad'' \cite{Hui:2005} and
presents some ``similarities'' when observed in different environments
(see \cite{Barrat:2015} for a short review). 

\corr{However, those comparisons were performed on data taken
in some similar sociological environments, which are typically
occidental, educated and often with a scientific background
(in conferences or high-school). 
Here we wish to extend the study of face-to-face interactions by comparing them 
to some very different datasets that were originally designed for other aims.
The fist one are the data taken in the rural Malawi village. The
second one concerns interactions among baboons in a primatology
center.}

\corr{Moreover, there is more information in the data than what was previously
presented \cite{Cattuto:2010,Barrat:2014} . Indeed, one has access to the full timeline of interactions
for \textit{each pair} of individuals separately (what we call in the following a
``relation''). This allows to study the mean-interaction time per
relation and, most importantly, deviations of the contact duration
from it, which reveals the underlying relation dynamics. 
We will show that they are surprisingly similar in all the settings.}

After describing our data selection and
methodological differences with some previous studies in \sect{method},
we will focus on the details of the temporal interactions in \sect{temporal} after
showing rapidly that social interactions among the participants are 
obviously very different in each environment.
We will introduce the concept of contrast of the contact
duration (deviation from the mean) and show that the distributions are
extremely similar on each dataset and for each relation individually.
In the Discussion part, we comment on the utility of using the robust
contrast distribution in improving agent-based models, \corr{and
  conclude summarizing the results and highlighting some possible future extensions.}
Some extra information, referred to in the text, is given in the \textit{Suplementary Information} (\SI) document.

\section{Material and methods}
\label{sec:method}

\subsection{Datasets}

We have chosen four datasets from the \textit{sociopatterns} web site,
sociologically most dissimilar.

\begin{enumerate}
\item \hosp : these are early data collected over 3 days \footnote{here
    and in the following, we will only consider complete (24 h) day periods.} on 75
  participants in the geriatric unit of a
  hospital in Lyon (France) \cite{Vanhems:2013}. Most
  interactions (75\%) involve nurses and patients.
\item \conf: these are also some early classical data from the ACM Hypertext 2009 (\safeurl{www.ht2009.org})
  conference  that involved about a hundred of participants
  for 3 days \cite{Isella:2010} in Torino (Italy). The audience is
  international with a scientific background.
  There exist also some data taken at another conference in Nice in 2009 (SFHH, \cite{Genois:2018})
  with more participants, but we prefer to use the former which has a
  number of individuals comparable to the other datasets. However we
  have checked that we obtain similar results with the SFHH data.
\item \malawi: these proximity data were taken in a small village of the district
  of Dowa in Malawi (Africa) where 86 participants agreed to participate for
  13 (complete) days. Interestingly those data contain both extra 
  and intra-household interactions, although we will
  not distinguish them here. This community consists
  essentially of farmers. 
\item \baboons Those data were taken at a CNRS Primate Center near Marseille
  (France) where 13 baboons were equipped with the sensors for a
  duration of 26 days. The goal was to study their interactions, and
  study how conclusions reached from data-analysis match those
  provided by human observation.
\end{enumerate}

With that choice, we span very different sociological environments.
We have also analyzed a few other datasets collected at the SFHH conference, an
office and a high-school. They give similar results (results are shown in the SI)
but we consider them as sociologically closer to the \conf one. 
We have chosen to focus on the  \textit{sociopatterns} data since they provide a consistent set 
taken with the very same devices, minimizing possible sources of systematic errors.

\subsection{Differences with previous studies}

Previous studies considered the overall temporal properties of
interactions, i.e. without differentiating the pair of people interacting.
In this work we will put accent on the temporal properties of each
pair separately.

Probability distribution functions (p.d.f) are often estimated by
histograms, i.e. by counting the number of samples that fall within
some bin. But for heavy-tailed distributions the size of the bins is
delicate to choose. 
With a constant size binning, several bins end up empty for large
values. Using a logarithmically increasing binning is neither a
solution since it supposes that the
distribution is constant on the wide range of last bins.
Following \cite{Newman:2005}, we will use instead the \textit{probability to
  exceed} function (\pte, also known as the ``complementary cumulative
distribution function'' or Zipf plot) which is computed simply by sorting the
samples and plotting them with respect to their relative frequency. In
this way, one does not need to define a binning and the distribution
is easier to apprehend.

\section{Results}

\subsection{Interactions between individuals}
\label{sec:struct}

\begin{figure}[htbp]
  \centering
  \subfigure[\hosp]{\includegraphics[width=0.49\linewidth]{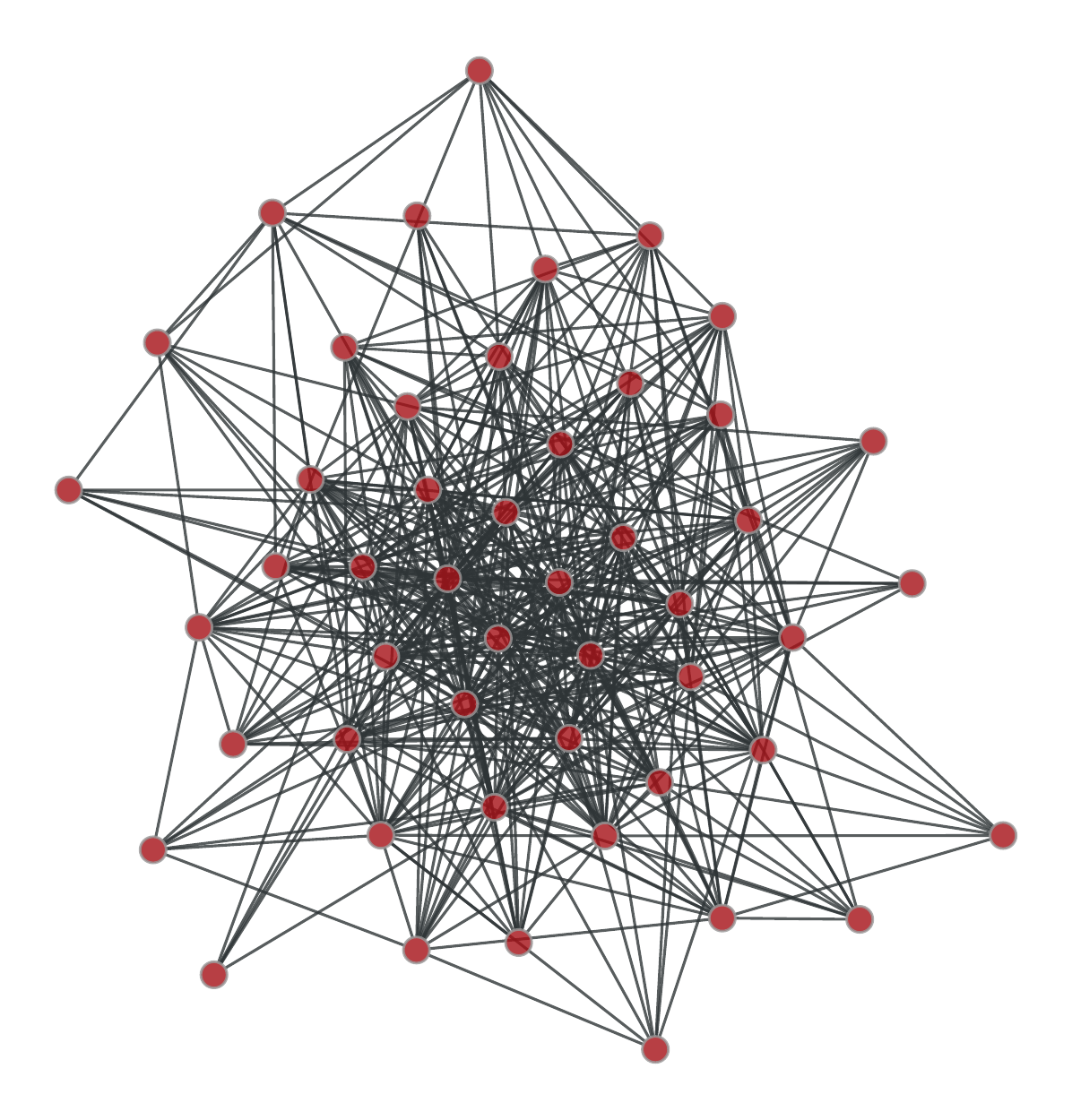}}
  \subfigure[\conf]{\includegraphics[width=0.49\linewidth]{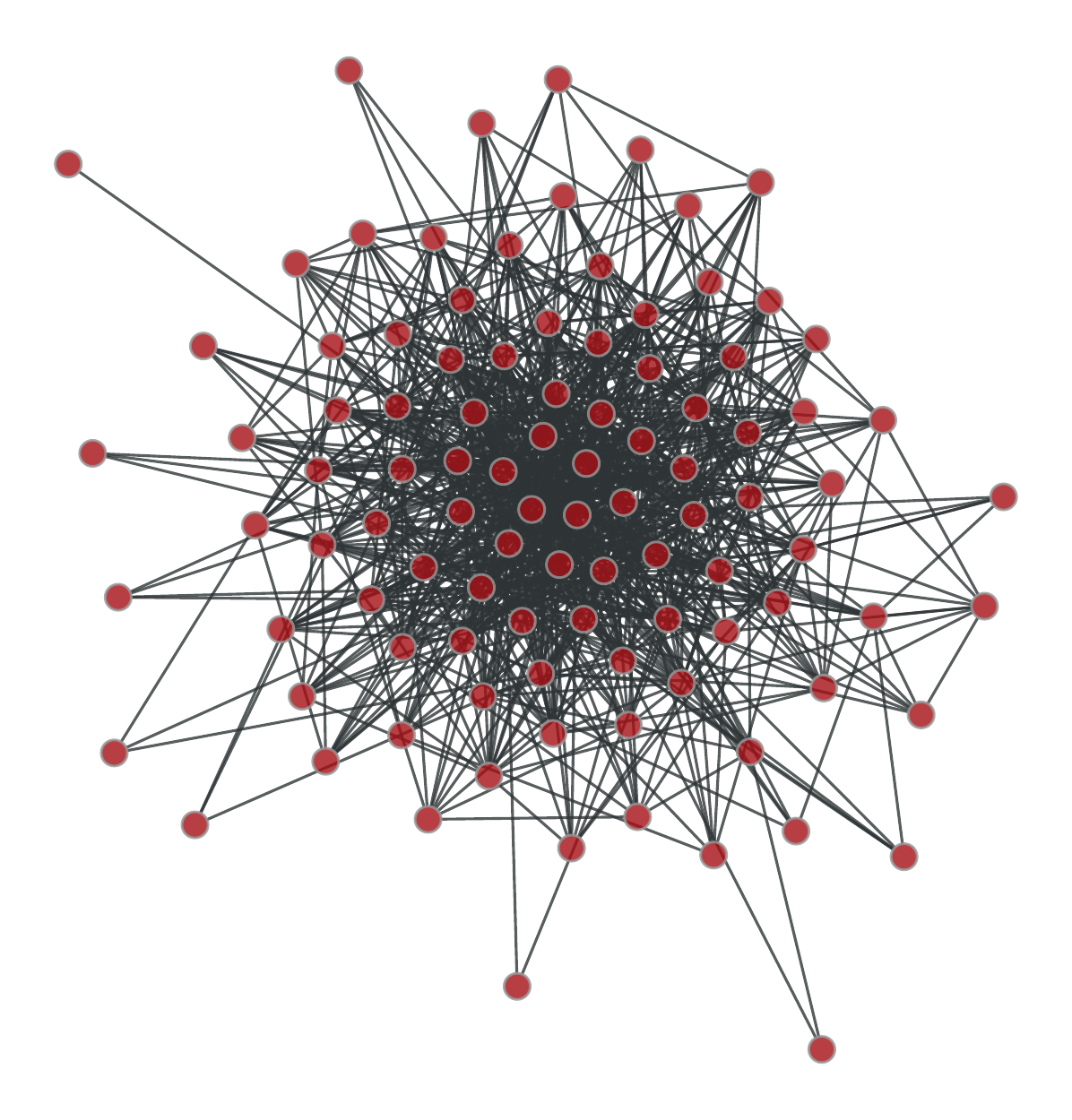}}\\
  \subfigure[\malawi]{\includegraphics[width=0.49\linewidth]{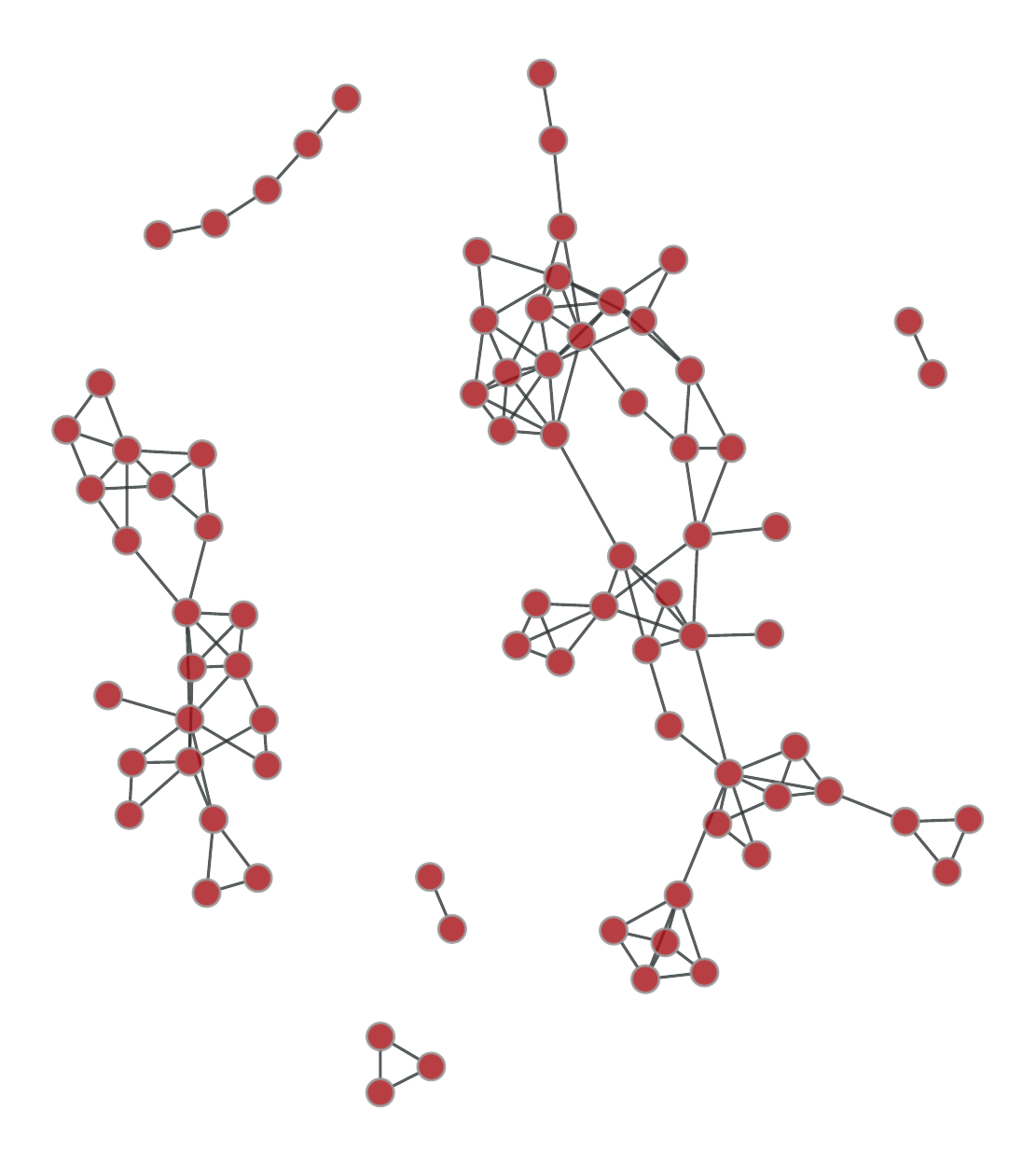}}
  \subfigure[\baboons]{\includegraphics[width=0.49\linewidth]{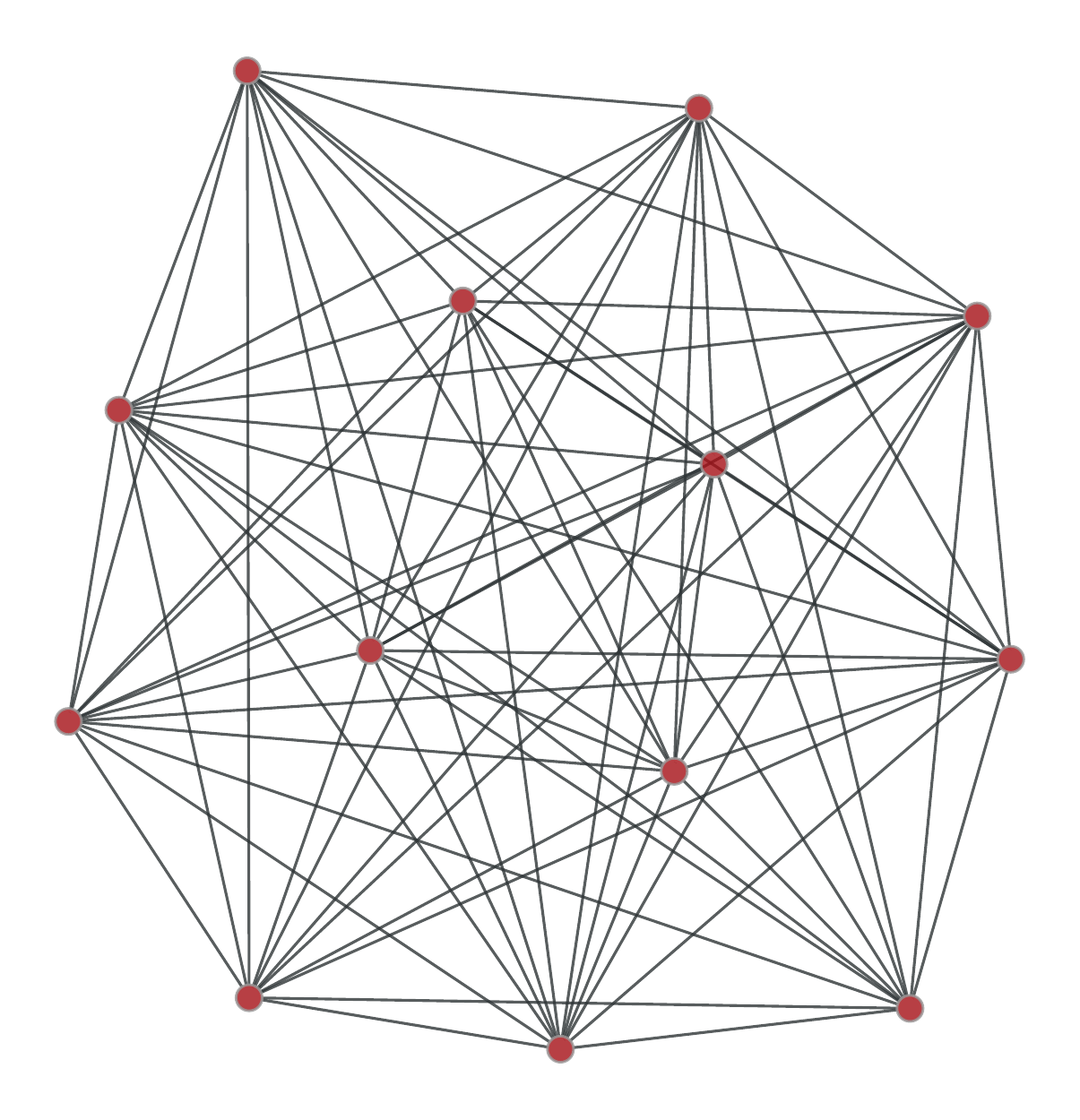}}
\caption{Aggregated graphs of interactions over one day for our 4
  datasets. Vertices (red points)
  represent agents and there is a link (edge) if there was at least
  one face-to-face interaction for more than 20 s. The first day
  from the datasets is used, but very similar results are obtained with
  the others.}
\label{fig:graphs}
\end{figure}

Since it is not our primary goal to study the social structures in those
very different communities, we just highlight visually some
differences on \Fig{graphs} which shows 24 hr time-aggregated graphs
of the relations between individuals.

The graphs for the \hosp and especially the \conf datasets show a strongly connected
core. The \malawi one is much sparser, while the \baboons one is
almost complete showing that each animal interact with all the others.

\begin{table}[htbp]
  \centering
  \begin{tabular}{lrrrrrrr}
  group & T &  N & \mean{k} & \mean{w}(mins) & \mean{s} (mins) \\
  \midrule      
  \hline
  \hosp & $3$ & $ 49\pm1 $ & $ 18\pm1 $ & $ 6\pm13 $ & $97\pm101$\\
  
  \conf & $3$ & $ 100\pm3 $ &  $ 20\pm1 $ & $ 2\pm14 $ & $ 46\pm63$\\
  
  \malawi & $12$ & $ 70\pm4 $ &  $ 3\pm1 $ &$24\pm37$ & $65\pm74$\\
  
  \baboons & $26$ & $ 13\pm1 $  & $ 11\pm1 $ & $8\pm11$ &$87\pm53$\\
  \bottomrule
\end{tabular}

  \caption{\label{tab:graphs} Properties of time
    aggregated graphs on each dataset per day. Uncertainties 
    are the standard deviations between the days. 
    $T$ is the number of (complete) days in the dataset. $N$ the number
    of interacting agents.
    $\mean{k}$ is the mean degree, i.e. the average number of agents each
    individual interacts with during one day. 
    $\mean{w}$ is the mean weight where the weights
    specify the total duration of a single relation \cite{Barrat:2004}.  
    Mean strength \mean{s}
    which represents the average total interaction time per individual.
  }
  \end{table}

\Table{graphs} gives a more quantitative view of some of the graph's properties.
The number of different people met per day (the degree of the graphs) is
about 20 in both the hospital and the conference environments.
As is apparent in \Fig{graphs}(c), it is is much smaller in the rural
community (3). 
But the interaction times are longer (\about 25 min) which reflect
different sector of activities (agrarian and including inter-housing
relations for the \malawi data). 

The strength of the relation represents the total time per
individual spent interacting
with others per day. It is essentially the product of the mean number
of people met per day by the
time spent interacting with them ($\mean{s}\simeq\mean{k}\mean{w}$).
It varies
by a factor of two (from 45 min to 1.5 h) although the large
standard-deviations indicates important daily variations due to the 
heavy-tail of the distribution.

The comparison to the \baboons dataset should be handled with care
since there is a much smaller number of agents (13).
Since each baboon interacts essentially with each other
(\Fig{graphs}(d)), the mean degree is bounded to $\mean{k}\simeq N$.
On the other hand, their small number possibly increases their
interaction duration  
($\mean{w}$) so that the strength of their relation is finally similar to
that of the human groups.

The goal of this short section is not to dwell into the topological details of these
time evolving graphs, 
but to illustrate that, as expected, these heterogeneous sociological groups
show some very distinct interaction patterns between individuals.

\subsection{Face to face temporal relations}
\label{sec:temporal}

\begin{figure}[ht!]
  \centering
  \includegraphics[width=\linewidth]{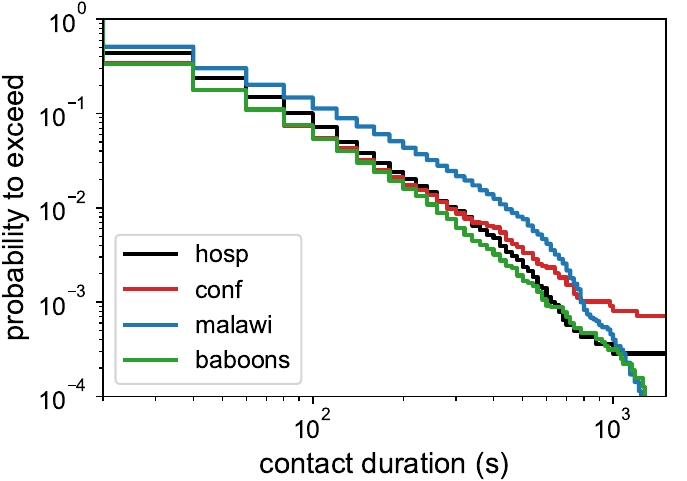}
  \caption{\label{fig:tij} Distribution (\pte) of the contact
    duration on the four datasets (all days used). There is one entry
    for each contact of each pair of individuals so that both aspects
    are inter-mixed.
}
\end{figure}

We are interested in the duration of the contacts in those different networks.
\Fig{tij} shows a classical distribution, that of the duration of
contacts. We emphasize that such a
representation mixes all the interactions of all the participants in the 
same plot.
As well known, these distributions are ``heavy-tailed''; most
interactions are of short duration  (at the minute level) but some may drift
up to an hour.
Interactions for people in \malawi tend to last longer than for
all the others.
The baboons' duration of interaction
is similar to the human ones (as noticed in \cite{Gelardi:2020}), although there
are some sizable differences at short times, somewhat squeezed by the
logarithmic scale.
Overall, although there is a common trend, some differences appear too.

\corr{The new aspect of this work concerns the detail of each relation
  separately. }
For a given data-taking period, each relation consists in a set of intervals measuring the beginning and
end times of the interaction at the resolution of the instruments (20 s). 
There is a varying number of interactions (intervals) per relation, 
that we call $\Ncon(r)$.
In the following we will consider the duration of the
interactions
that we note $\{t_i(r)\}_{i=1,\cdots,\Ncon(r)}$. They are thus
variable-size timelines expressed in units of the resolution step.

\begin{figure}[ht!]
  \centering
  \subfigure[]{\includegraphics[width=\linewidth]{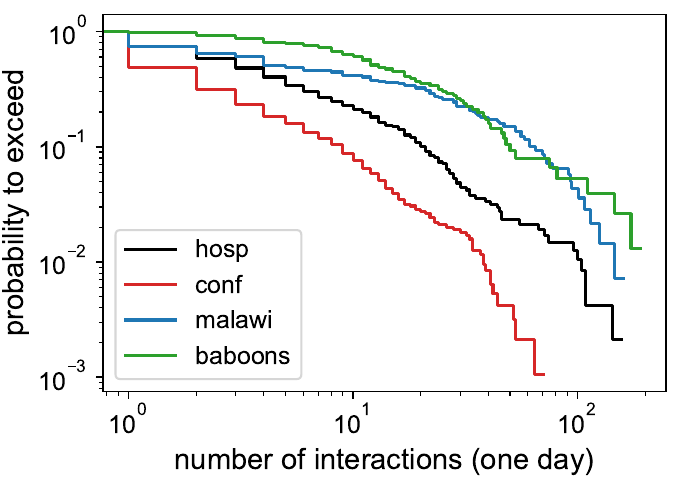}}
  \subfigure[]{\includegraphics[width=\linewidth]{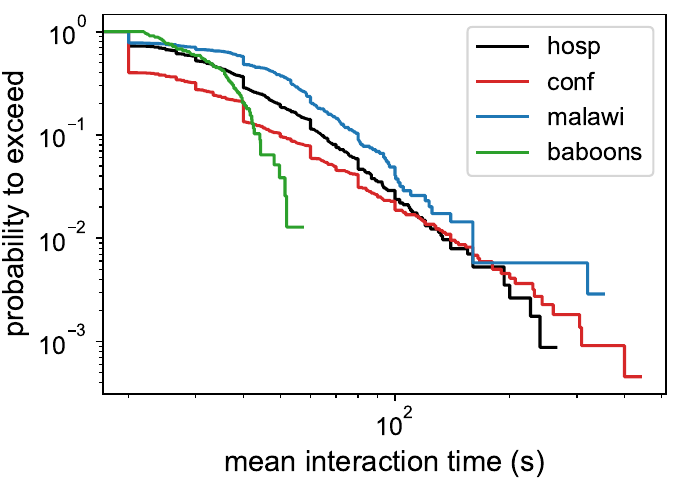}}
  \caption{\label{fig:Ncon} General characteristics of temporal
    relations on the 4 datasets. (a) Distribution (\pte) of the number of
    interactions per relation for one day, and (b) of the mean interaction
    time. To gain precision, we use the complete datasets for the latter.}
\end{figure}

The number of registered interactions for a given pair depends on the
total duration of the experiments (Table \ref{tab:graphs}) but we may
compare them just for one day.
The distribution of this variable is shown in \Fig{Ncon}(a). 
It is clearly different for each group. 
People at the conference tend to interact (with the same person) less
often. In 65\% of the cases it is only once per day, against 25\% for the \hosp and
\malawi datasets, and 3\% for \baboons.

The mean interaction time per relation
\begin{align}
\label{eq:meantr}
  \bar t(r)=\dfrac{1}{\Ncon(r)}\displaystyle{\sum_{i=1}^{\Ncon(r)} t_i(r)},
\end{align}
is shown in \Fig{Ncon}(b). 
Here again distributions are heavy-tailed and different. There is a
marked difference between animals and humans, the former interacting
for shorter times.

\begin{figure}[ht!]
  \centering
  \subfigure[]{\includegraphics[width=\linewidth]{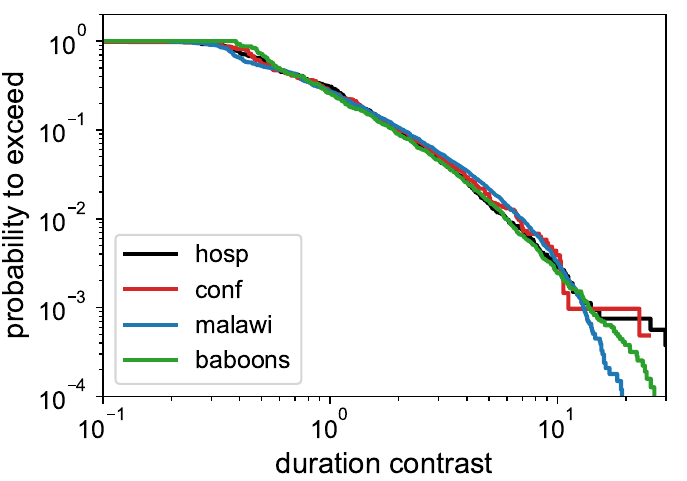}}
  \subfigure[]{\includegraphics[width=\linewidth]{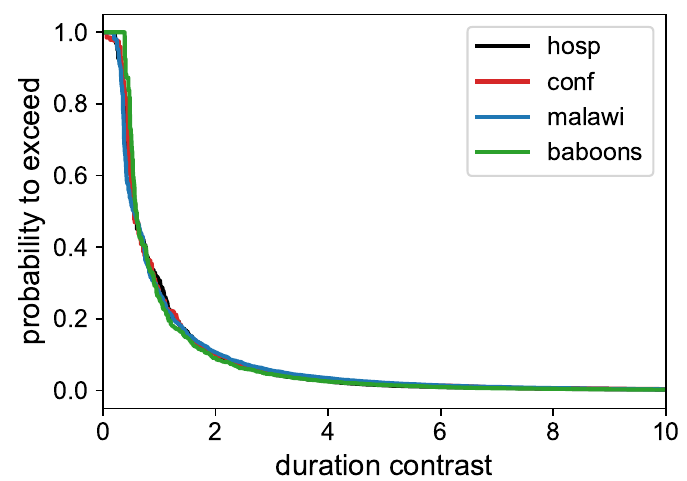}}
  \caption{\label{fig:cdc_all} Distributions (\pte) of the duration
    contrast obtained for all relations  within the same group 
    satisfying $\Ncon(r)>50$ in logarithmic (a) and linear scales (b).
    \corr{The complete datasets have been used (i.e all days).
      The interaction mean-time corresponds to a value of 1. We
    then see for instance that the probability for an interaction to
    last longer than its mean-time is around 30\%, but, rarely, it can
    exceed 10 times the mean-time}.}
\end{figure}

\corr{
We are now interested in studying the \textit{deviations} of the contact duration 
from the mean value for a given relation. Indeed, in physics the
dynamics of a process is often revealed by such a quantity.
For instance in cosmology, one uses the ``density contrast'' that
represents the galactic
density divided by its mean value. It is the fundamental quantity
which traces the dynamics of the underlying field (see e.g.
\cite{Peebles:1980}).
Inspired by this example, we propose to study what we call the
```duration contrast'', or simply ``contrast'' which is the simplest dimensionless
quantity we can form to study deviations from the mean-value}

\begin{align}
  \delta_i(r)=\dfrac{t_i(r)}{\bar t(r)}, 
\end{align}
\corr{
where $r$ recalls that the quantity varies for each relation.}
The contrast represents our tendency to spend more or less time than
usual with a given individual. Note that ``usual'' is
meant as the mean-interaction time between the two peculiar agents (\Fig{Ncon})
and varies for each relation.
For a small number of samples, the arithmetic mean (\refeq{meantr}) is however a 
poor estimate of the true mean-time and also strongly correlated
to the individual samples. Taking the ratio leads to a very noisy
estimate of the true contrast variable.
In the following we will then apply a cut to keep timelines with a
sufficient number of samples. Since the distributions are very broad we
require at least $\Ncon(r)>50$ contacts in a relation. We will study
later the effect of this cut on the results.
On the complete datasets, we are left with respectively 
57, 26, 91 and 70 timelines for the \hosp, \conf, \malawi and \baboons
datasets.
We show the \pte distributions of the contact duration contrast for
the 4 groups in \Fig{cdc_all}.
 
\corr{The tails look now  very similar up to 10 times the mean-time.}
The same distribution is observed on data from another conference, an office
and a high-school (\Sup{S2}). Thus, a (very) similar distribution is
observed on 7 independent datasets.

\corr{
To be more quantitative and assess the level of compatibility between the distributions, we
use a Monte-Carlo method. For each dataset, we numerically invert the
empirical distribution functions (which are one minus the \pte's shown
on \Fig{cdc_all}) to construct the inverse cumulative function $F^{-1}$. We
then draw $N$ numbers $u$ from a $[0,1]$ uniform distribution, transform
them with $F^{-1}(u)$ and reconstruct the \pte. The procedure is repeated 100
times and all distributions are plotted on top of each other on \Fig{mcerror}.

\begin{figure}[ht!]
  \centering
  \includegraphics[width=\linewidth]{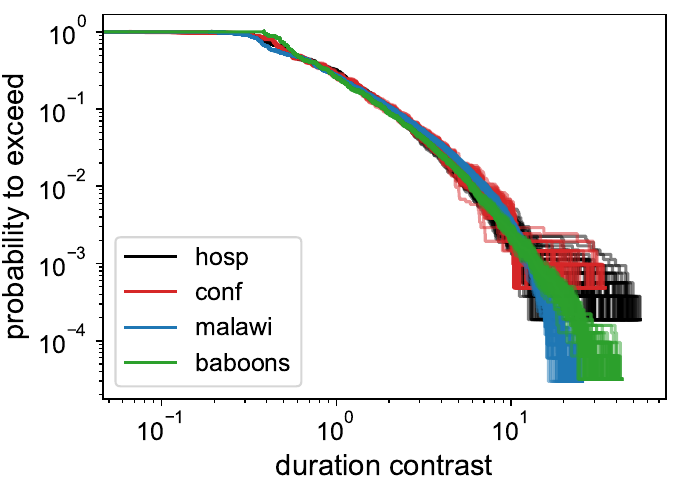}
  \caption{\label{fig:mcerror} Distributions (\pte) of the duration
    contrast obtained with the Monte-Carlo method described in the
    text to estimate numerically the statistical spread for each
    dataset. Each color represents a possible realization of the same dataset.}
\end{figure}

One sees that the distributions are indeed all compatible in the
$0.6\lesssim\delta\lesssim10$ range, where the upper bound comes from the
limited sample size of the \hosp and \conf datasets, and the
lower one from slight (but statistically significant) differences for low values.
This will be our range of interest in the following.
}

\corr{Since the data-taking periods are very heterogeneous (ranging
  from 3 days for the \conf and \hosp datasets, to 12 and 26 for the
  \malawi and \baboons ones respectively) we have split the data day
  by day and verified that no particular one(s) particularly affects
  the results (\Sup{S3-1}). We have also removed randomly a fraction of
  the agents (up to 50\%), i.e. we removed all relations involving those
  agents, which did not affect the contrast
  distributions in a sizable way (\Sup{S3-2}. Both tests confirm the robustness of the result.
}

\corr{ 
Another option for studying deviations from the mean is to use the $z$-score
\begin{align}
  z_i(r)=\dfrac{t_i(r)-\bar t(r)}{\sigma(r)},
\end{align}
where $\sigma$ represents the standard-deviation of the duration
values.
The results obtained with this variable are very similar to the
ones with the contrast (\Sup{S4}) and we did not notice any
difference on the tests that are presented later. Since the contrast variable is somewhat
simpler (the $z$-score involving second order statistics) we only focus in the following on it.
}

\begin{figure}[ht!]
  \centering
  \includegraphics[width=\linewidth]{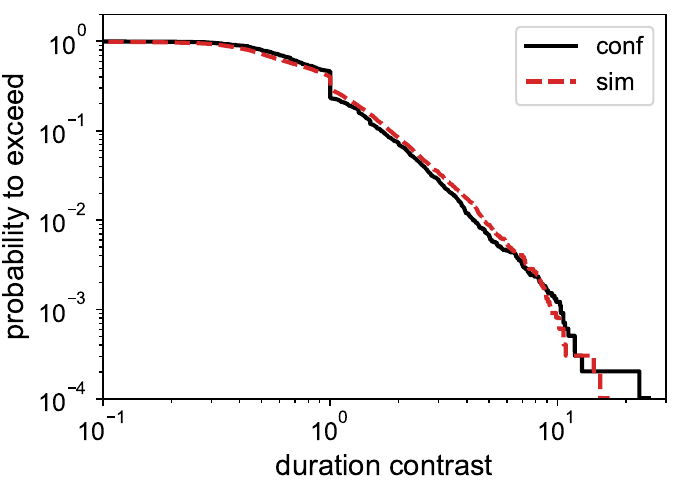}
  \caption{\label{fig:conf_sim} Distributions (\pte) of the duration
    contrast obtained for all relations in the
  \conf dataset and simulations produced using the corresponding
  \Fig{cdc_all} distribution (see text for details). The dip at 1
  comes from numerous cases (65\%) where \Ncon=1 always leads to $\delta=1$.}

\end{figure}

We consider the impact of applying the $\Ncon(r)>50$ cut.
First, we note that similar results are
obtained with a lower cut value as $\Ncon>30$ (\Sup{S5}). 
We then show that we can still reproduce the contrast distribution 
without any cut, using only the distributions with the cut (\Fig{cdc_all}).
To this purpose we perform Monte-Carlo simulations. For a given
dataset, for each relation (without any cut), we draw $\Ncon(r)$ random numbers 
following \Fig{cdc_all} distribution to obtain
$\delta_{i=1,\cdots,\Ncon}$ contrast values.
Those samples are obtained from the distribution with the
$\Ncon(r)>50$ cut, so with precise mean values that we call $\mu$.
We may mimic the statistical fluctuations due to any $\Ncon(r)$ value,
by using the ratio
 \begin{align}
  \delta_i^{mes}=\dfrac{\delta_i}{\tfrac{1}{\Ncon}\sum_i \delta_i}
   =\dfrac{t_i/\mu}{\tfrac{1}{\Ncon} \sum_i t_i/\mu}
   =\dfrac{t_i}{\bar t}
\end{align}
since $\mu$ actually cancels out. 
We compare the measured contrast distribution to the one observed on data, this
time without any \Ncon(r) cut, in \Fig{conf_sim} for the \conf
dataset. We reproduce correctly the whole contrast distribution
using only the \Fig{cdc_all} one obtained with $\simeq$1\% of the data ($\Ncon>50$).
Similar results are obtained on the other datasets (\Sup{S6.1}).
This shows that the contrast distribution obtained from the large sample
statistics is sufficient to reproduce any number of interactions,
including small-sample ones.
In other words, the $\Ncon(r)>50$ cut only cleans the data 
without affecting the underlying ``true'' contrast distribution.

\begin{figure}[ht!]
  \centering
  \includegraphics[width=\linewidth]{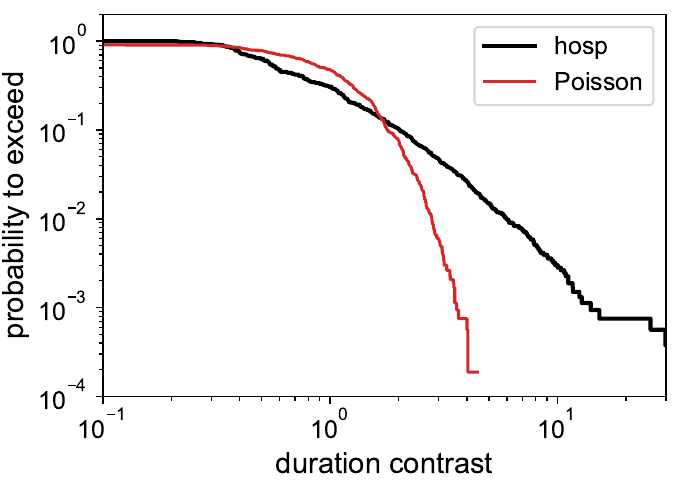}
\caption{\label{fig:Poisson} Distribution of the contrast duration
  assuming a Poisson distributed duration (in red). The
  parameters are taken from the \hosp
  dataset. The result obtained in real data is shown in black.}
\end{figure}

To check that the contrast distribution is not artificially produced
by the procedure of dividing the timelines by their mean value,
we use the \hosp dataset to retrieve the set of interacting agents and
their corresponding characteristics $\Ncon(r)$ and $\bar t(r)$.
We then draw $\Ncon(r)$
random numbers following a Poisson distribution of parameter $\bar
t(r)$ and recompute the contrast. The result is shown in \Fig{Poisson}
which is clearly different from the results observed on the data.

The shape of the observed contrast distribution (\Fig{cdc_all}) is nontrivial. It is
neither of exponential nor of power-law form. A stretched-exponential form is neither satisfactory.
\corr{Empirically, we could obtain
a reasonable fit in the $0.6 \lesssim \delta \lesssim 10$ region, by
combining both a power-law and an exponential function}
\begin{align}
\label{eq:fit}
  p(>\delta)=0.3 e^{-0.2\delta}/\delta^{1.1}
\end{align}

\corr{The denominator is here to enhance short contrasts, while the
exponential term describes the long ones. This could be an indication
of the existence of two regimes, one for short times when communications are more
informative and a longer one when real conversations form \cite{Button:2022}
}

\begin{figure}[ht!]
  \centering
  \subfigure[\hosp]{\includegraphics[width=0.46\linewidth]{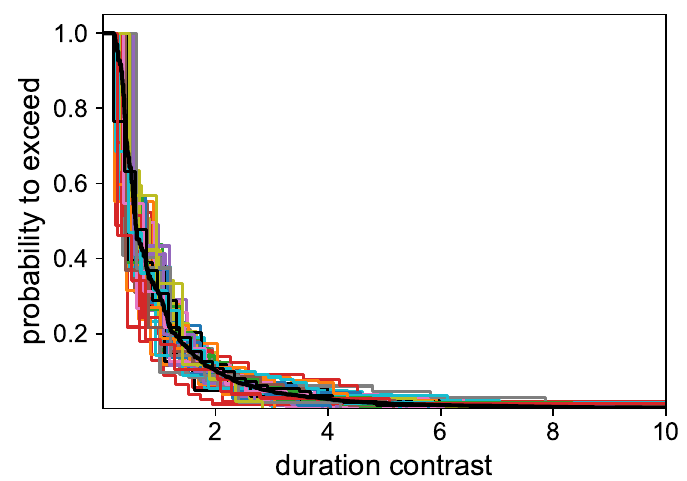}}
  \subfigure[\conf]{\includegraphics[width=0.46\linewidth]{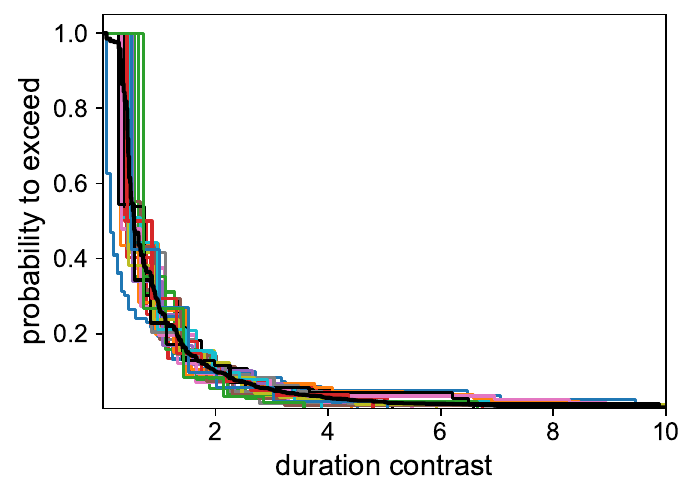}}
  \subfigure[\malawi]{\includegraphics[width=0.46\linewidth]{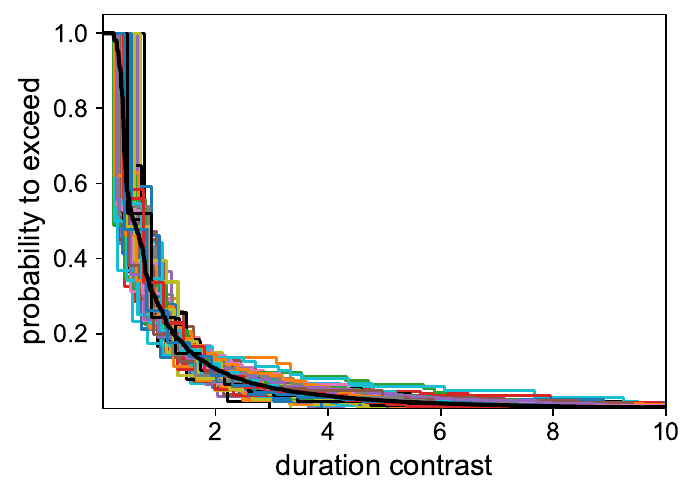}}
  \subfigure[\baboons]{\includegraphics[width=0.46\linewidth]{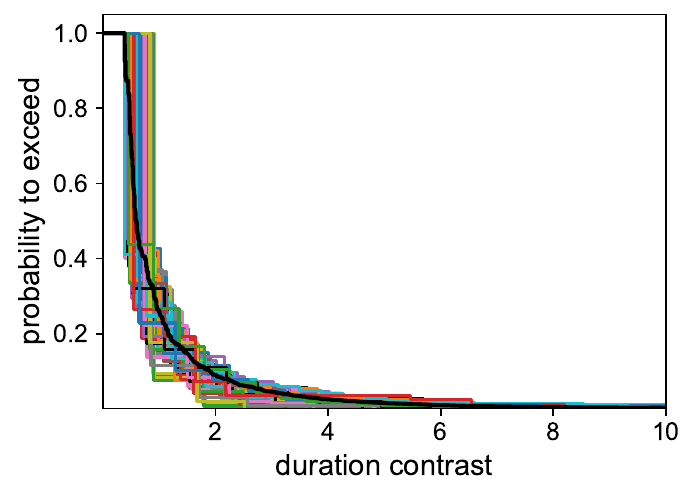}}
\caption{\label{fig:each_cdc} Distributions (\pte) of the contact duration contrast for
  each relation with at least 50 contacts. Each color represent a
  different distribution. The black line is the combined \pte shown 
  in \Fig{cdc_all}.}
\label{fig:cdc}
\end{figure}

At this point, we have shown that the \textit{combined} contrast
duration (i.e. for all relations) follows a very similar distribution.
We now consider \textit{each} relation separately and show in \Fig{each_cdc} a
superposition of the contrast duration distributions with the
$\Ncon(r)>50$ cut (similar results are observed without it but are, 
as expected,  more noisy (see \Sup{S6.2}).

They \textit{all} follow rather closely the common contrast distribution.
In other words, while
the choice of individuals we meet (\Fig{graphs}), the interaction rate (\Fig{Ncon}(a)) and
mean-time spent together (\Fig{Ncon}(b)) varies strongly with the environment,
the propensity to spend more (or less) time than usual 
with a given individual, is remarkably similar. This points to the
idea that once a face-to-face contact is triggered it follows its own
dynamics, out of the sociological context.


For the sake of completeness, we note that we found no sizable
correlations between the contact duration within the timelines (see \Sup{S7}). This
indicates one can draw independent samples using \refeq{fit}.

\begin{figure}[h!]
  \centering
  \includegraphics[width=\linewidth]{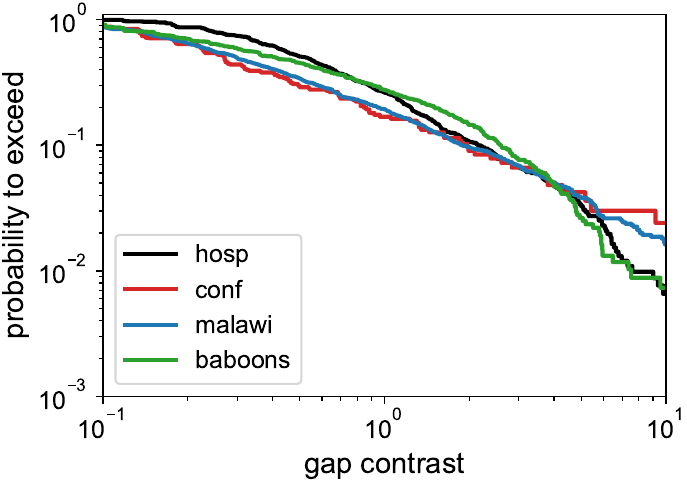}
  \caption{Distributions of the contrasts of the gap-time
    (inter-contact duration) on our datasets. 
    To avoid the long night breaks, we show results for a single day.}
  \label{fig:gap}
\end{figure}

We also considered the inter-contact (or ``gap'') time in the
relations to see whether its contrast reveals features similar to the duration ones. This is not the case as
shown in \Fig{gap}. The contrast of the inter-contact time thus seems to be
more dependent on the sociological context.

\section{Comparison with a model}
\label{sec:fdm}

\begin{figure}[h!]
  \centering
  \includegraphics[width=\linewidth]{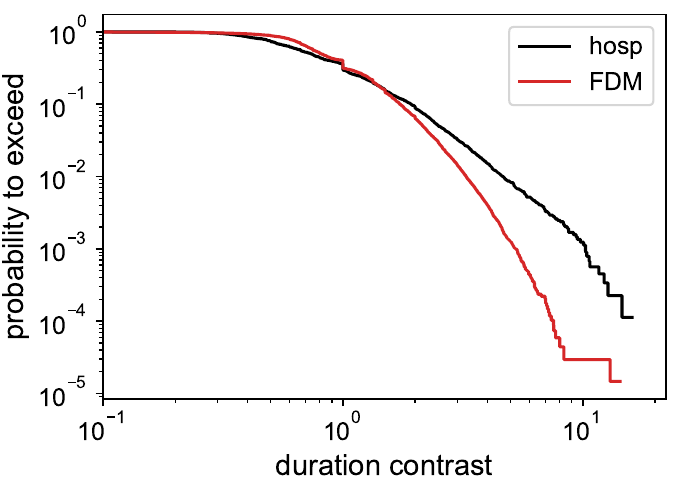}
  \caption{Comparison of the contrast distributions obtained
    with the \hosp dataset to the result of the ``force-directed
    motion'' (FDM) model \cite{Flores:2018}.
    We used the parameters provided by the authors and
    their dataset (slightly different from ours, dues to a different selection).
    The FDM curve is the combined result from 10 simulations.}
  \label{fig:fdm}
\end{figure}

The contrast distribution can be used as a new metric when studying face-to-face
temporal graphs in order 
to test and improve existing agent-based models designed to reproduce the
full evolution of a set of individuals.
For instance, the ``force directed motion'' (FDM) model is successful in
describing several key features of observed face-to-face interactions
\cite{Flores:2018}.
Based on the idea of attractiveness between some agents performing a
random-walk within a bounded perimeter \cite{Starnini:2013,Starnini:2016}, the model further
includes the concept  of ``similarity''  between two individuals
\cite{Papadopoulos:2012}, 
known as homophily in social sciences. The similarity $s_{ij}$ 
influences the time two agents spend together and the way the random-walk is biased.
The model assumes that the contact duration between two agents
is exponentially distributed with a rate $s_{ij}/\mu_1$, where $\mu_1$
is adjusted on the data to reproduce the overall duration of contacts.
We have run the code provided by the authors
with their setup corresponding to the \hosp dataset, to test the distribution of
the contrast variable. \Fig{fdm} shows that the model distribution falls too steeply.
We have tried adapting the parameters and some parts of the code but
could not find a configuration giving a better contrast distribution (see \Sup{S8})
\footnote{The authors of \cite{Flores:2018} also quote some results
obtained with a hyperbolic geometry \cite{Krioukov:2010} but, due to
the lack of public software, we could not test it}.

Modeling correctly the tails of the contact duration is also essential
in epidemiological studies since the spread of a disease happens mostly during long
interactions. For a given mean-interaction time, \refeq{fit} allows to
simulate a much more realistic duration of contacts than a Poissonnian one.
This can be used in SIR-like statistical inference, or using
agent-based models, for the precise modeling of long interactions.

\section*{Conclusion}
\label{sec:discussion}

We have compared face-to-face interaction data taken in some very 
different environments; some were recorded in a European hospital and 
during a scientific conference, others in a small village in Africa.
\corr{
With the original intention to pinpoint differences with the results
concerning humans, we have also included
data on baboons' interactions in an enclosure.} 

\corr{
Although the topological structures (who interacts with whom) and the mean-time spent
together are clearly dependent on the sociological environment, it
appears that the deviations from the mean-time for each pair (do we
spend more/less time than usual with a given person) follow a very
similar distribution, including for baboons.
We (and baboons) tend to interact most often for much less time than
``usual'' with a given individual and sometimes, but rarely, much longer. 
What is striking is that the distribution for this quantity, which we
call the ``relation contrast''
looks universal. It is the same for people at a scientific conference or
farmers in a small Malawi village (and baboons in an enclosure), 
see \Fig{cdc_all} (also \Sup{S2} for the 7 datasets).}

These results suggests that, once a face-to-face contact is
triggered, it follows its own dynamics independently from the social context.
This is maybe not a big surprise to a sociologist in particular
working in the field of Conversation Analysis \cite{Button:2022} where it is
postulated that each conversation follows some rules independently from
the social context \footnote{The fact that similar results are
  observed on baboons requires however to enlarge the concept of ``discussion''.}.
But to our knowledge, this was not noticed by physicists \corr{and may
help disentangling the topological and temporal aspects of face-to-face interactions.}

\corr{The possible universality of the relation contrast must be challenged with more data.
On the animal side, one should consider groups of animals with strong
social interactions, that can be identified (labeled) and followed
individually. Hominids, as baboons, are known to have social behaviors
close to ours, which probably explains the similarity of the
contrast distribution with the human's one. Chimpanzee or bonobo's data
should show similar characteristic.
Concerning mammals, we could think of tracking individuals in elephant herds or wolf packs
but it's difficult to acquire precise data in the wild.
The most promising approach concerns the study of social insect
networks \cite{Fewell:2003}. Details about ant interactions is probably the most
feasible since recent techniques allow to tag and follow each
individual separately \cite{Greenwald:2015}.
On the human side, we need to check whether the contrast is influenced by age.
Since children perceive time differently from adults, following the contact patterns
of young children in a nursery could provide a valuable insight into
this question.
}


\begin{backmatter}


\section*{Availability of data and materials}
\begin{itemize}
\item The datasets analyzed during the current study are available in the \textsl{sociopatterns} repository,  \safeurl{www.sociopatterns.org}
\item The FDM code was downloaded on 10 June 2023 from \safeurl{https://bitbucket.org/mrodrflr/similarity\_forces}
\item The \textsf{python3} software used to produce the results is available from 
\safeurl{https://gitlab.in2p3.fr/plaszczy/coll}
\item  The graph-related computations and \Fig{graphs} were obtained with 
the \textsf{graph-tool} (\textsf{v 2.43}) software \cite{Graph-tool:2014}. 

\end{itemize}


\bibliographystyle{bmc-mathphys} 

\bibliography{references}

\end{backmatter}

\end{document}


\bigskip
\title{ 
On the duration of face-to-face contacts
\\\textbf{Supplementary Information}}
\author{ S. Plaszczynski, G. Nakamura, B. Grammaticos, and M. Badoual}
\maketitle
\tableofcontents
\newpage

\section{Main result}
\label{sec:Main}

Our main result concerns the duration of face-to-face contacts between
\textit{each} pair of individuals. Although
the mean-interaction time varies, the \textit{deviation} from this mean value is very similar in several very different social 
contexts. As in other fields of physics (e..g cosmology) we call this
over/under-duration the duration \textit{contrast}. It is a dimensionless value
that describes how much given an interaction is longer or shorter than its
usual (mean) time and can be expressed  in percents. By construction its mean value is 1.

To remove some statistical noise , we only use data where there is a sufficient number
of interactions (samples) \Ncon. Since the duration distribution is heavy-tailed we
use a large minimal value of 50.

\begin{figure}[ht!]
  \centering
  \subfigure[]{\includegraphics[width=0.49\textwidth]{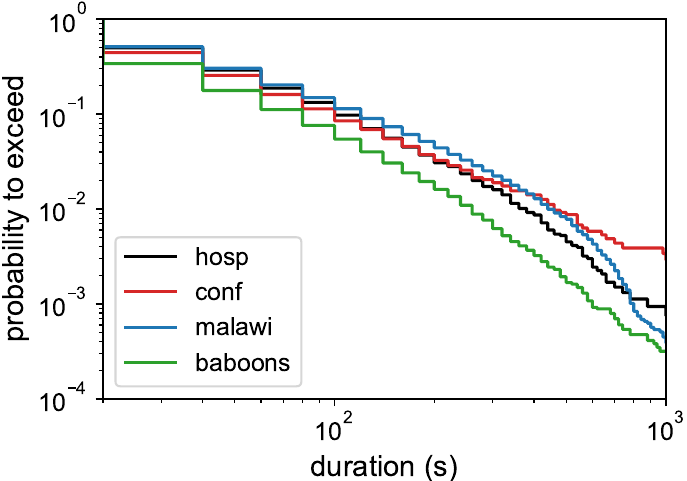}}
  \subfigure[]{\includegraphics[width=0.49\textwidth]{cdc_all_log}}
\caption{\label{fig:cdc} (a) distributions (\pte) of the duration of contact and (b) contrast
duration on the 4 datasets. The cut $\Ncon>50$ is used for both.}
\end{figure}

\Fig{cdc} (a) shows how the distribution of the duration of contacts
and how it changes when computing the contrast in (b). Both use the
$\Ncon>50$ cut.

\section{Other datasets}
Although in our work we have focused on sociologically 
most dissimilar datasets, we have also looked at some other data
provided by the \textit{sociopattern} collaboration.
\begin{enumerate}
\item \sfhh: these are data taken at another conference (SFHH,\cite{Cattuto:2010,Stehle:2011})
with \about 380 participants for 2 days.
\item \office:  data taken at an office (Intitut de Veille Sanitaire
  near Paris,
  \cite{Genois:2015,Genois:2018}) with about 165 participants for 10
  days.
\item \highschool: data from a french high-school, near Marseille  \cite{Fournet:2014,Mastrandrea:2015}.
  About 300 participants for 4 days 
\end{enumerate}

\begin{figure}[ht!]
  \centering
  \subfigure[]{\includegraphics[width=0.49\textwidth]{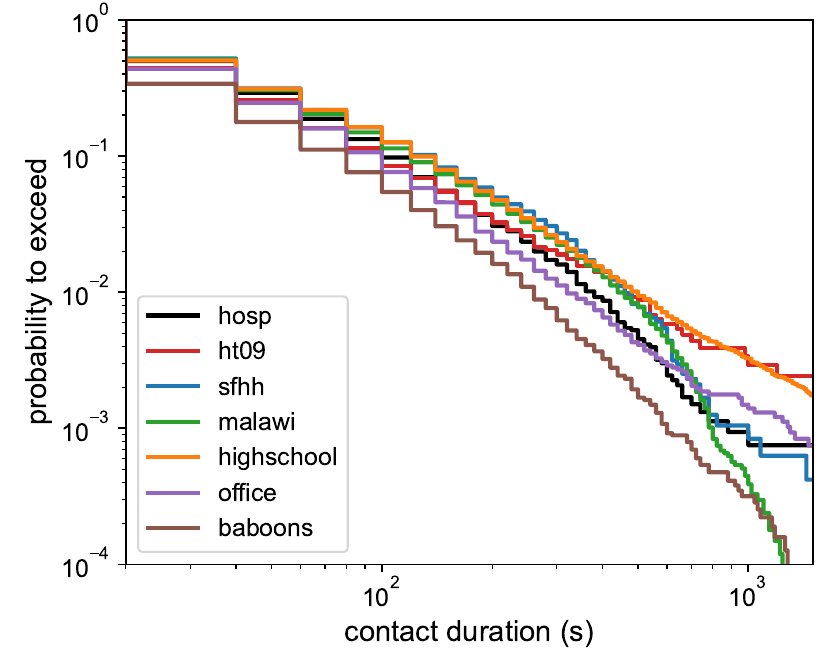}}
  \subfigure[]{\includegraphics[width=0.49\textwidth]{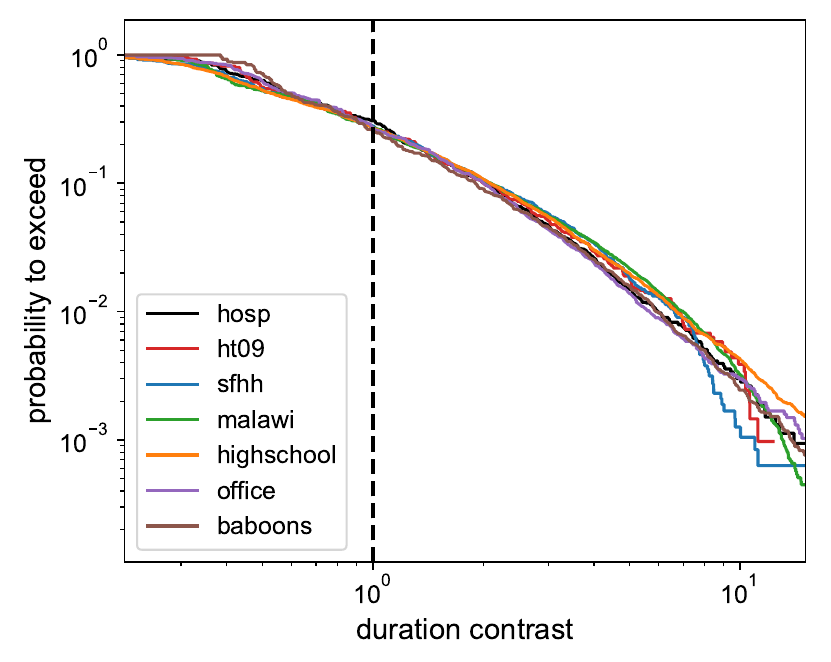}}
\caption{\label{fig:cdc_new} (a) distributions (\pte) of the duration of contact and (b) contrast
duration for 3 other datasets.}
\end{figure}

As in \sect{Main} we show how using the contrast standardizes the
contact duration on \Fig{cdc_new} on these new datasets; for comparison we
also included the previous \conf result and still use the $\Ncon>50$ cut.

The contrast distributions are similar to the results presented in
the paper, and differences are barely noticeable in a linear representation.

\begin{figure}[ht!]
  \centering
  \subfigure[]{\includegraphics[width=.7\textwidth]{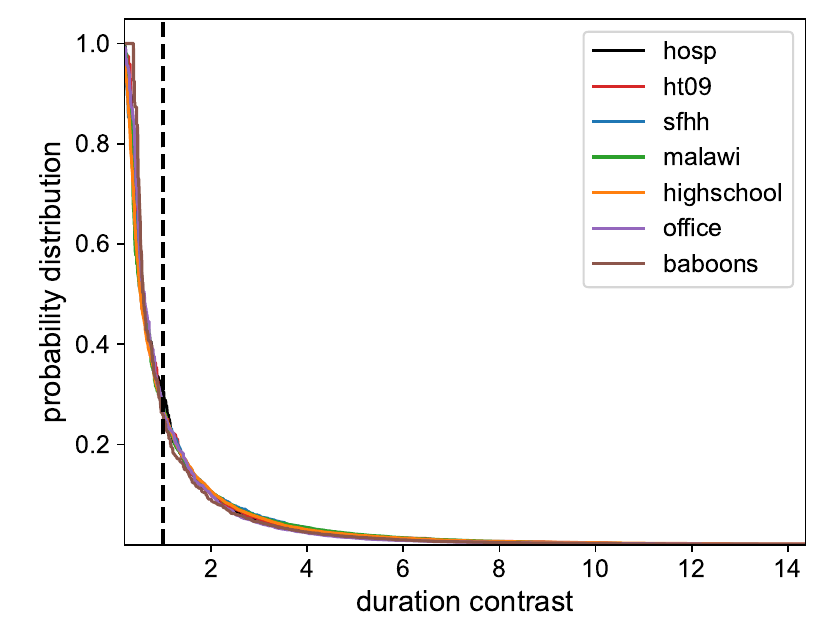}}
\caption{\label{fig:cdc_new_lin} Same as \Fig{cdc_new} (b) in linear scale.}
\end{figure}

\section{On the robustness of the results}

\subsection{Day by day variation}
\begin{figure}[ht!]
  \centering
  \subfigure[]{\includegraphics[width=0.49\textwidth]{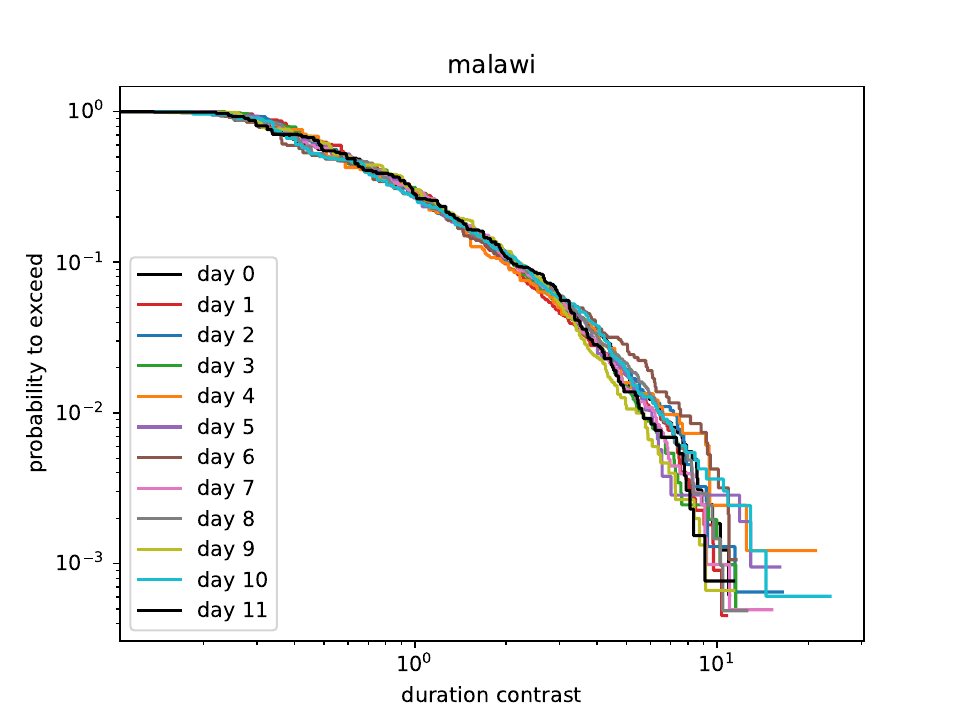}}
  \subfigure[]{\includegraphics[width=0.49\textwidth]{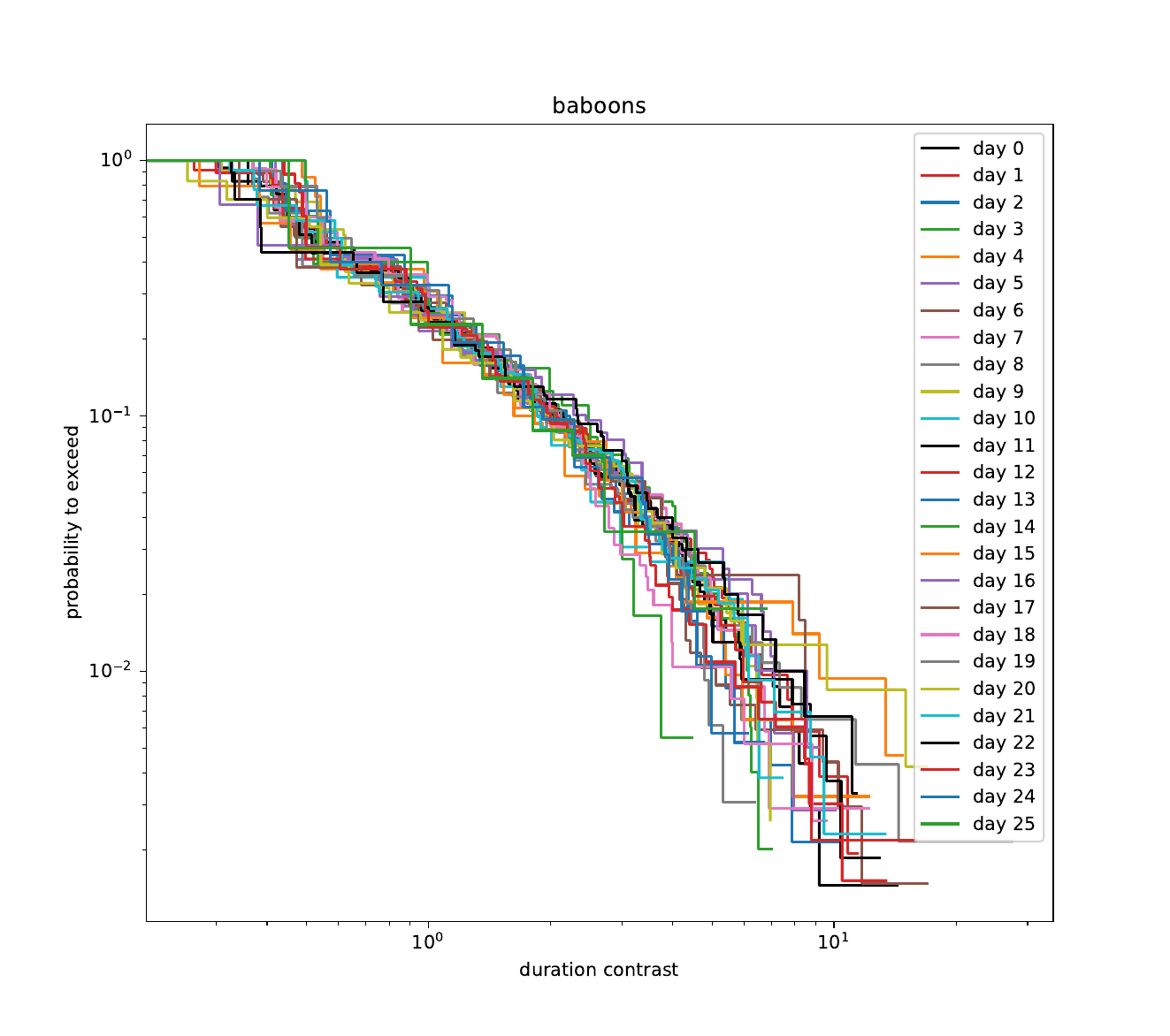}}
\caption{\label{fig:days} Distributions (p.t.e) of the duration
  contrast for the \malawi (a) and \baboons (b) datasets splitting the
samples day by day.}
\end{figure}

The datasets are very heterogeneous in size as shown in the paper Table 1. 
The \malawi and \baboons ones have more data-taking days (12 and 26). 
Increasing the data size actually affects essentially the very end of
the p.t.e tail as 
shown on the paper Fig.5, since it probes very rare events.
To check that a single day does not bias the results for the lengthy
datasets, we have split the data day by day and the results are presented
on \Fig{days}.

\subsection{Varying the network size (and topology)}

The number of agents (then size of the graphs) varies among
the datasets (see Table 1).
As shown on Fig 1., the topologies are also very different. 
If there is a common process different sizes
and topologies of the network were thus tested. 

\begin{figure}[ht!]
  \centering
  \includegraphics[width=0.55\textwidth]{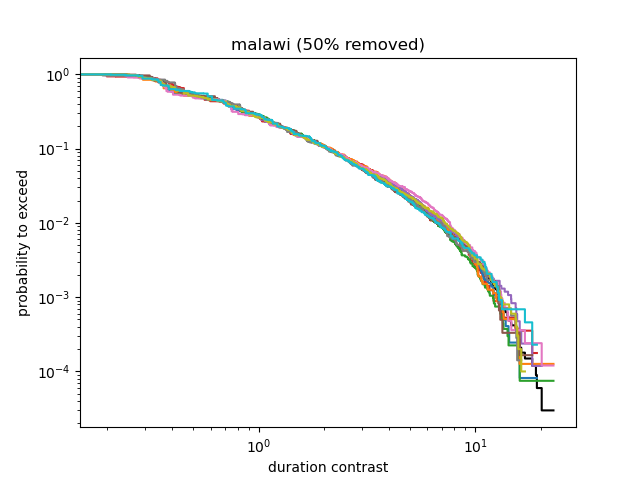}
\caption{\label{fig:malawi_rm} Distributions (p.t.e) of the duration
  contrast for the \malawi datasets after removing 50\% of the agents.
Different colors represent different random realizations of the
removed agents.}
\end{figure}

To test whether a particular set of individuals could influence the
shape of the contrast distributions
we have further performed the following test: we removed some fraction
of random individuals from the full list of relations. The distributions
obtained when removing (randomly) 50\% of individuals on the \malawi
dataset are shown on \Fig{malawi_rm} (similar results are obtained
with the other datasets).

It shows that the distributions are very robust to the size of the network (and thus
its topology) and not driven by a particular set of vertices.

\section{z-scores}
\label{sec:z}

We have characterized deviations from mean-values by the contrast
variable which is simply obtained by rescaling the sample values by
the mean.
Another dimensionless quantity that can be build to probe deviations
from mean values is the z-score

\begin{align}
  z_i(r)=\dfrac{t_i(r)-\bar t(r)}{\sigma(r)},
\end{align}
where $\sigma$ represents the standard-deviation of the values.

\begin{figure}[ht!]
  \centering
  \subfigure[]{\includegraphics[width=0.49\textwidth]{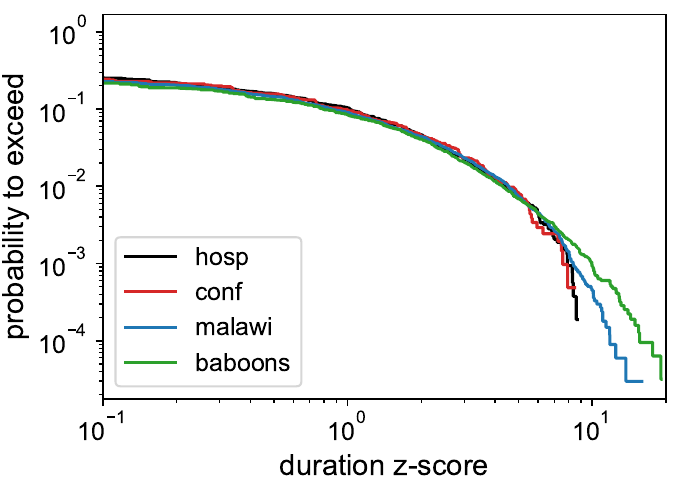}}
  \subfigure[]{\includegraphics[width=0.49\textwidth]{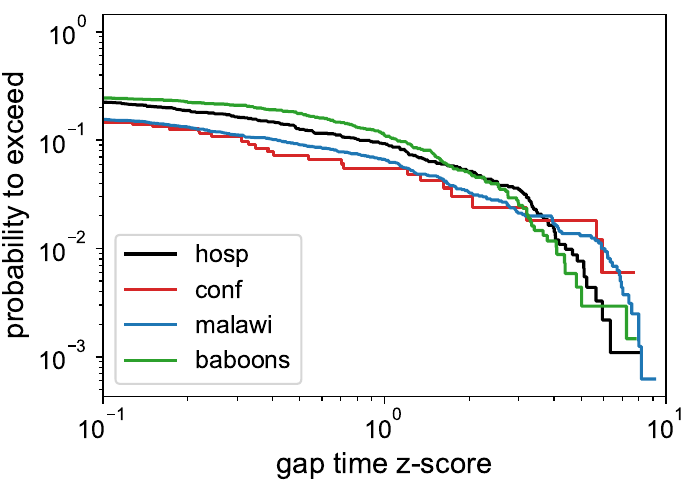}}
\caption{\label{fig:zscore} (a) Distributions (\pte) of z-scores for
  the duration (b) and gap-time. Note that the values can
  become negative and cannot be shown on such logarithmic plots (which
  is why distributions do not start at 1).}
\end{figure}

Here $t_i$ is the contact duration or the gap time.
We show the p.t.e distributions for both quantities on \Fig{zscore}.

\begin{figure}[ht!]
  \centering
  \includegraphics[width=0.49\textwidth]{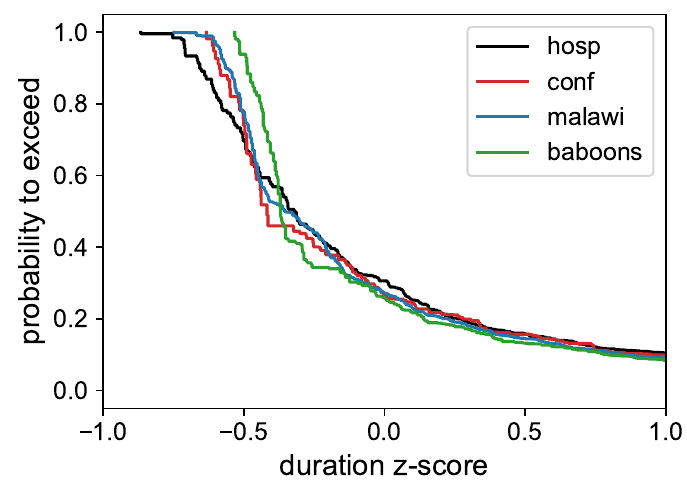}
\caption{\label{fig:zscore_zoom} Zoom of \Fig{zscore} on the negative part
  of the distribution using a linear axis.}
\end{figure}

The distributions show a behavior that is similar to the one obtained using the
contrast variable. They may even seem more regular at small
z-scores, but this is an artifact due to using a logarithmic axis.
Indeed z-scores can become negative (while the
contrast cannot). Representing it for the duration with a linear scale
one obtains \Fig{zscore_zoom}.
There are some differences around -0.5 which are the counterpart of 
what is observed near
0.4 on the contrast plot (\Fig{cdc} (b)).

Using the z-score thus does not change our conclusion on the gap-time which
looks more context-dependent.
Overall the results obtained with z-scores are very consistent with
the ones obtained with the contrast on which we focus in the paper.

\section{Number of interactions}

\begin{figure}[ht!]
  \centering
  \includegraphics[width=0.7\textwidth]{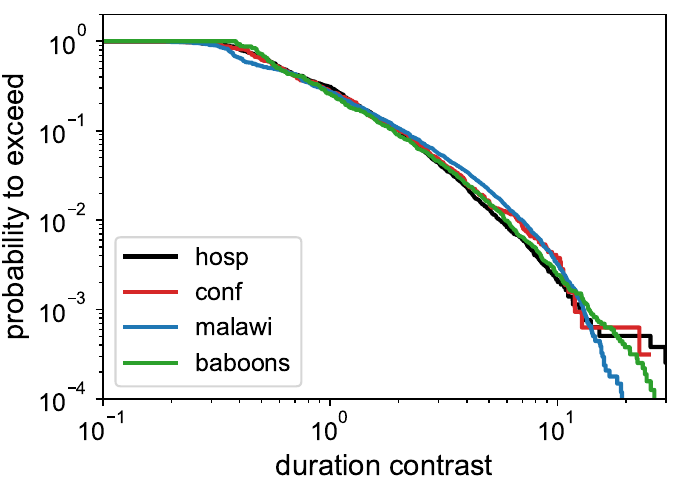}
\caption{\label{fig:Nint30} Distribution of the contrast duration on
  the 4 datasets  using a $\Nint(r)>30$ cut.}
\end{figure}

To compute a mean value one traditionally invokes the central limit
theorem and compute a reliable arithmetic mean with $\simeq 20$
samples . However the contact duration distribution is very wide
(\Fig{cdc}(a)) so we prefer to increase that value to $\Ncon>50$
before computing the contrast, still keeping a reasonable number of
timelines in each case.
We may use a lower cutoff and the contrasts distributions are still
similar(\Fig{Nint30}) . But we have added some noisy samples. Note that
this value of 50 is only dictated by the data limit data-taking period.
Had we a really long period, the mean interaction times between each
pair would be well known and this cut would be not necessary.

\section{Simulations without a minimal number of interactions}

\subsection{Combined contrast}

\begin{figure}[ht!]
  \centering
  \subfigure[\hosp]{\includegraphics[width=0.49\textwidth]{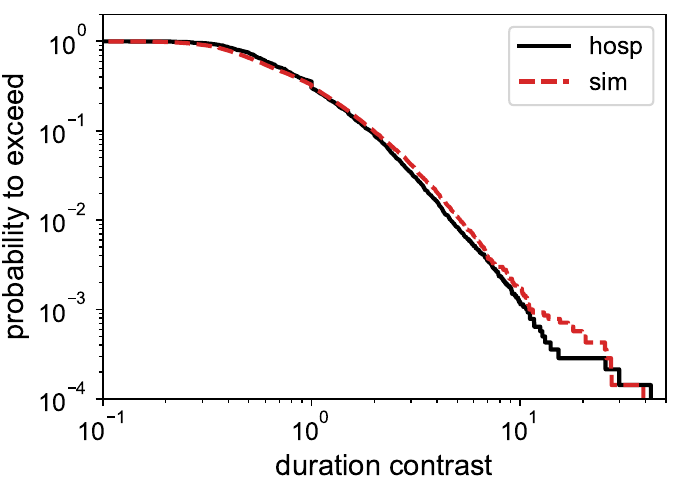}}
  \subfigure[\conf]{\includegraphics[width=0.49\textwidth]{conf_sim}}
  \subfigure[\malawi]{\includegraphics[width=0.49\textwidth]{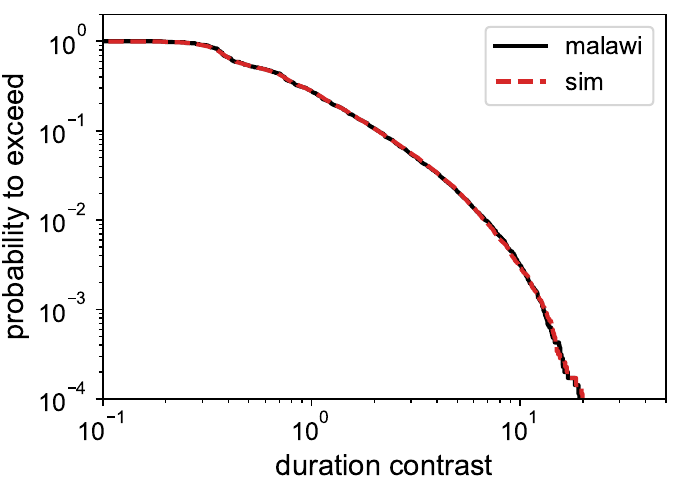}}
  \subfigure[\baboons]{\includegraphics[width=0.49\textwidth]{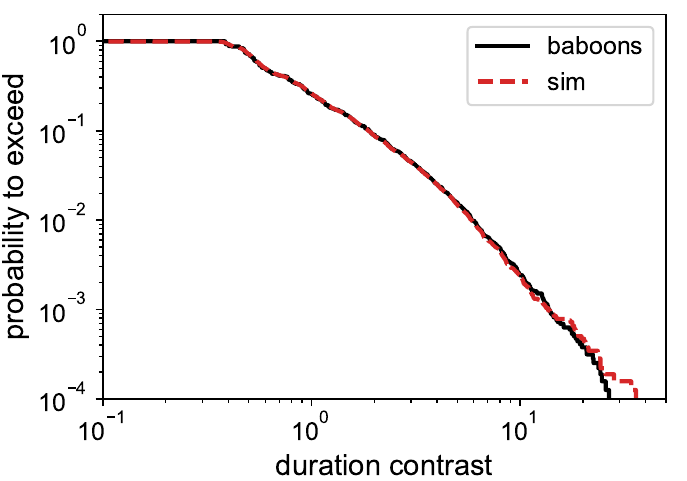}}
\caption{\label{fig:mc} Results of the simulations described in the
  manuscript (Sect 3) without using a cut on \Ncon for all the datasets.}
\end{figure}

We have shown in the manuscript how to reproduce the noise present on
the data for all $\Nint$ values on the \conf dataset in Section 3.
\Fig{mc} shows the results for the other datasets. We
reproduce all the data, using only a small
fraction of the timelines (the ones with $\Ncon>50$ which represents
respectively ).

\subsection{Contrast per relation}

In the previous section, the duration (and contrast) of all contacts are combined on
\Fig{mc} in the sense the contrast of each relation are mixed together.
We now consider each relation and show its contrast \pte with
different colors. The timelines are noisy since we do not use any
minimal number of steps (\Nint). The \baboons result is less noisy but this
is only due to the fact that the timelines have more samples due to
the longer data-taking period (26 days).

\begin{figure}[ht!]
  \centering
  \subfigure[\hosp]{\includegraphics[width=0.49\textwidth]{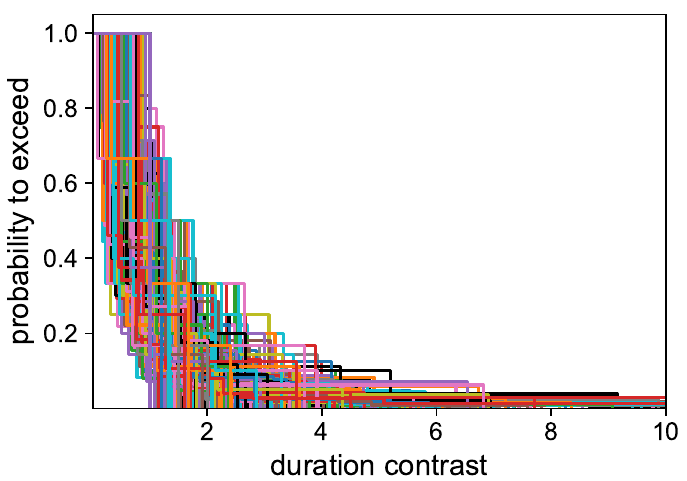}}
  \subfigure[\conf]{\includegraphics[width=0.49\textwidth]{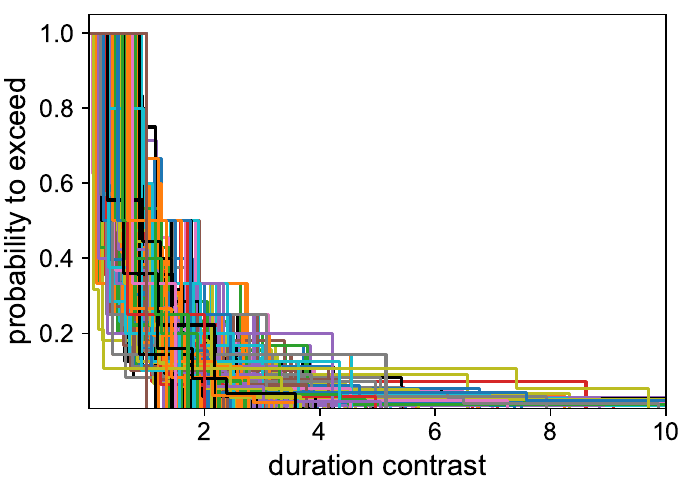}}
  \subfigure[\malawi]{\includegraphics[width=0.49\textwidth]{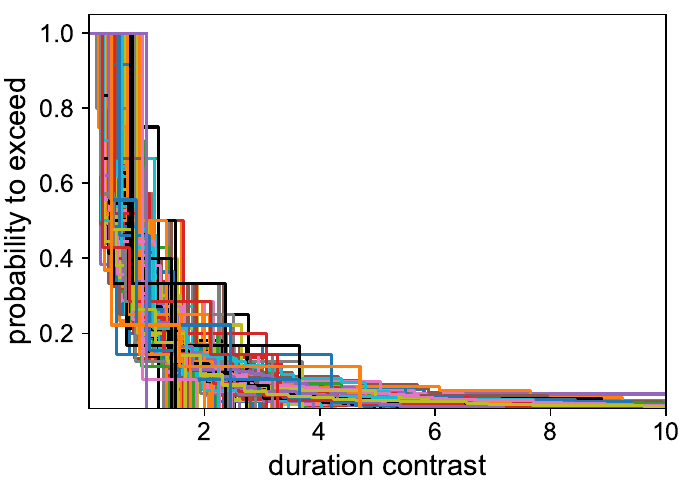}}
  \subfigure[\baboons]{\includegraphics[width=0.49\textwidth]{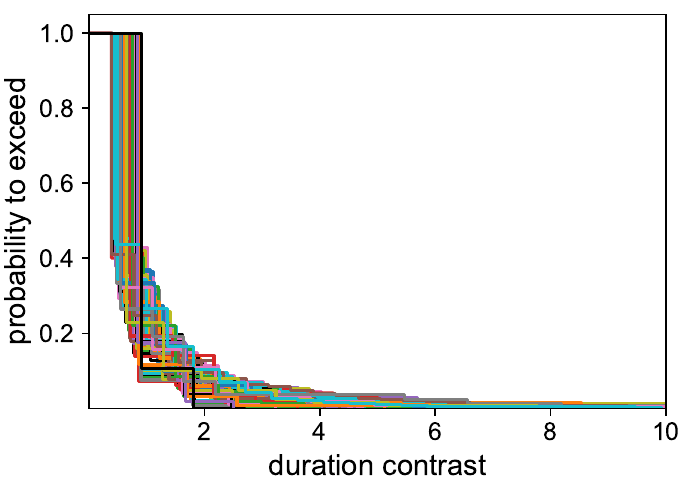}}
\caption{\label{fig:cdc_all} Contrast distributions per relation
  measured on the data without any \Nint cut. Each color represents a different
  relation timeline.}
\end{figure}

\begin{figure}[ht!]
  \centering
  \subfigure[\hosp]{\includegraphics[width=0.49\textwidth]{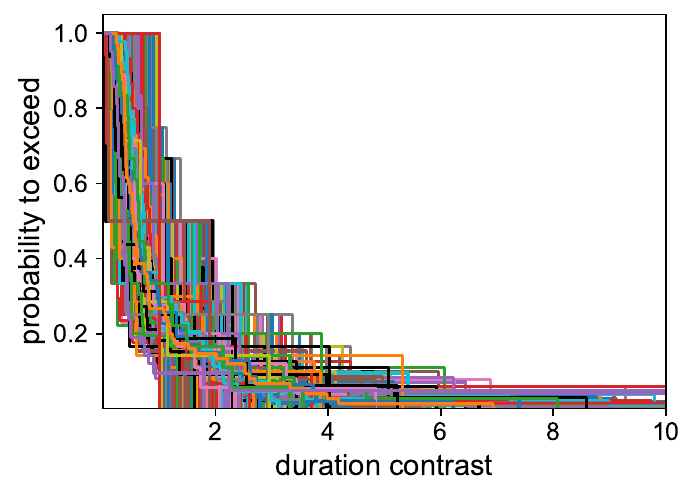}}
  \subfigure[\conf]{\includegraphics[width=0.49\textwidth]{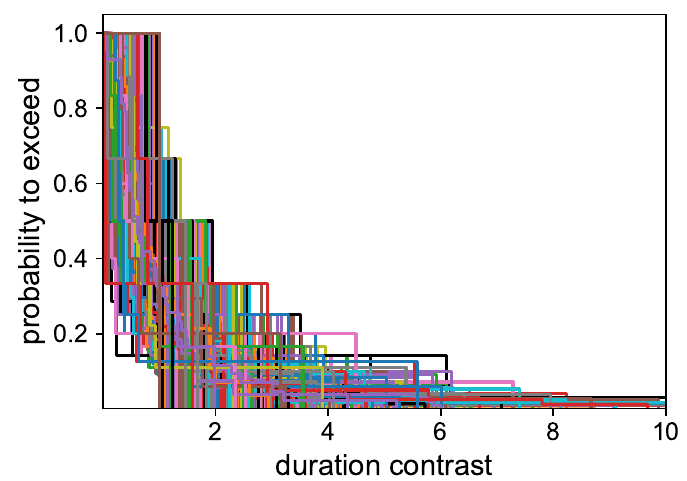}}
  \subfigure[\malawi]{\includegraphics[width=0.49\textwidth]{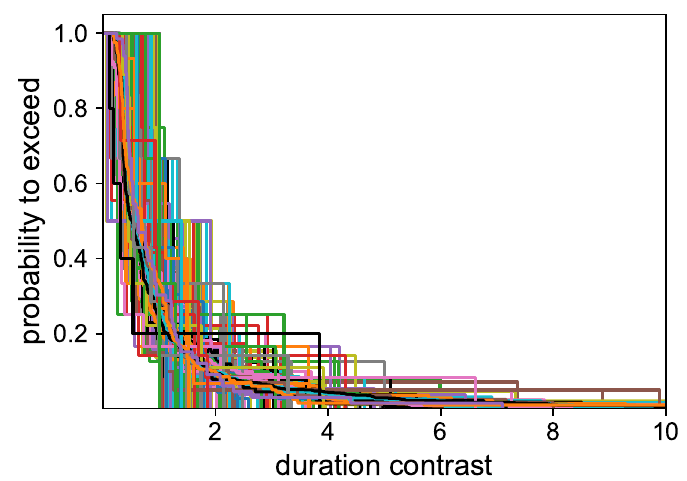}}
  \subfigure[\baboons]{\includegraphics[width=0.49\textwidth]{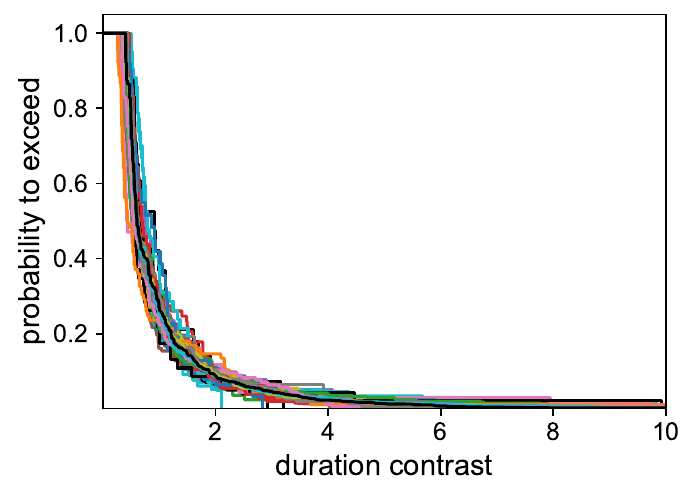}}
\caption{\label{fig:sim_all} Results of the simulations described in the
  manuscript (Sect 3) for each relation (colored curves) 
without using a cut on \Ncon for all the datasets.}
\end{figure}

We then perform the same simulations than described in the manuscript, but this
time on each relation individually. Once again only the global shape
of the contrast obtained with the $\Ncon>50$ cut (i.e. \Fig{cdc}(b)) is used to
draw the random numbers. The results are shown on \Fig{sim_all}. They
resembles closely the data (\Fig{cdc_all}). This confirms 
that the shape and spread observed on data (\Fig{cdc_all})
can be reproduced using only the cleaned \Fig{cdc} distribution
($\Ncon(r)>50$ and statistical fluctuations (on the mean) due to
the limited size statistics.

\section{Contact duration correlations}

\begin{figure}[ht!]
  \centering
  \includegraphics[width=\textwidth]{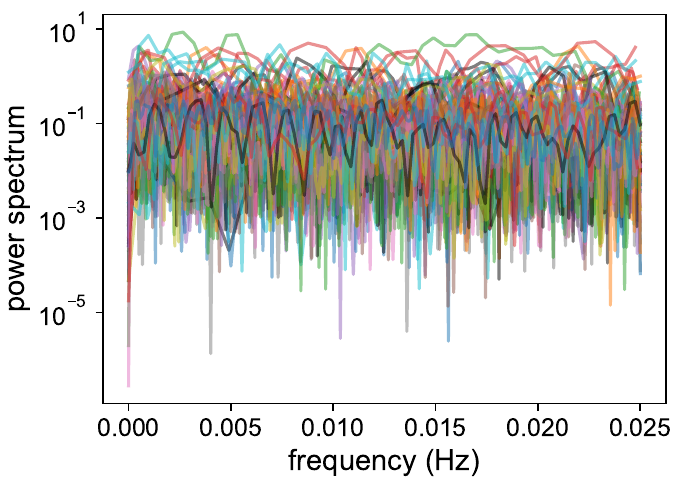}
\caption{\label{fig:psd} Power-spectra density measured }
\end{figure}

To test if some temporal correlations appear in the timelines, 
we compute the power-spectral density for each relation with the
classical Welch periodogram method. \Fig{psd} obtained with the \malawi data
shows the result.

The spectra are essentially flat indicating no sizable temporal
correlations within the timelines.

\section{FDM model}

The FDM model has in its base configuration 4 free parameters that
were adjusted by the authors of \cite{Flores:2018} to reproduce several
key aspects of the evolving temporal graph of relations.
The parameters are
\begin{enumerate}
\item $\mu_1$ which impacts the contact duration
\item $F_0,\mu_2$ that concern the way the random-walk is biased and
  impacts the appearance of recurrent communities
\item $L$ , the box size, tuned to reproduce the average degree.
\end{enumerate}

The contact duration contrast for each pair is defined by
\begin{align}
  \delta_i=t_i/\bar t,
\end{align}
so that the contact duration $t_i$ is a first essential ingredient.
The mean interaction time is
\begin{align}
  \bar t=w/\Ncon
\end{align}
so that the weight $w$ and the number of interaction \Ncon (between the 2
individuals) are also important.

In the following we have focused on the \hosp dataset provided by the authors.

We have run the model (10 simulations each time) with the base
parameters which are $\mu_1=0.8,\mu_2=0.9,F_0=0.12$.

We noticed that the tail of the contact duration does not match well
the data as shown in \Fig{fdm}(a). 
This is also apparent in the Supplementary Material of \cite{Flores:2018}
(Fig. 7(a)) but somewhat hidden by the logarithmic binning (see paper
Sect. I.B). We have looked for a better $\mu_1$ value and found that
$\mu_1=0.7$ gives a better agreement for the contact duration 
but rather in the intermediate values range. The other relevant distributions
\Fig{fdm}(b-d) are not very satisfactory and our adjustment
for $\mu_1$ has finally little impact on the resulting contrast distribution \Fig{fdm}(e).

\begin{figure}[ht!]
  \centering
  \subfigure[]{\includegraphics[width=0.49\textwidth]{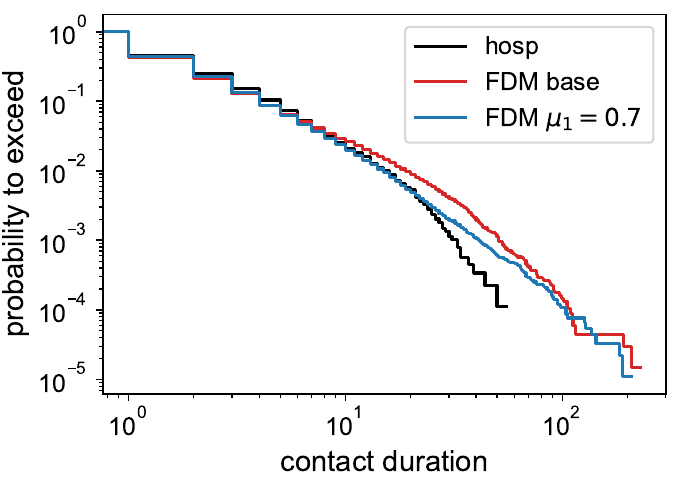}}
  \subfigure[]{\includegraphics[width=0.49\textwidth]{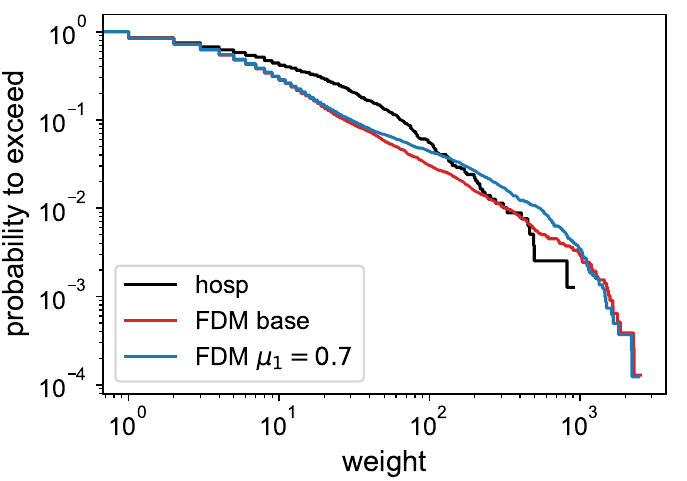}}
  \subfigure[]{\includegraphics[width=0.49\textwidth]{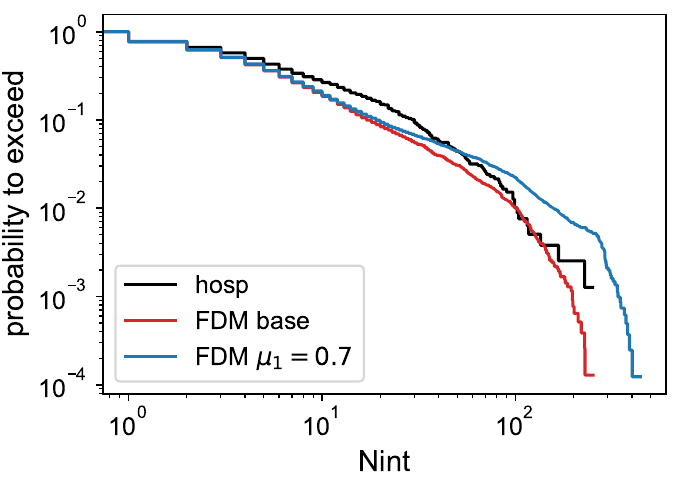}}
  \subfigure[]{\includegraphics[width=0.49\textwidth]{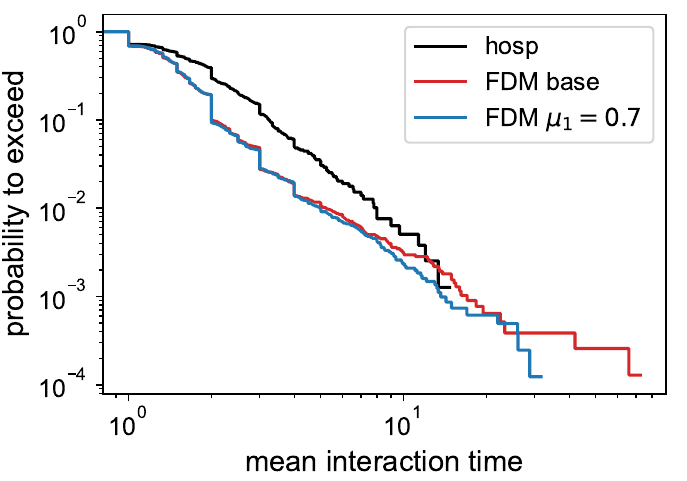}}
  \subfigure[]{\includegraphics[width=0.49\textwidth]{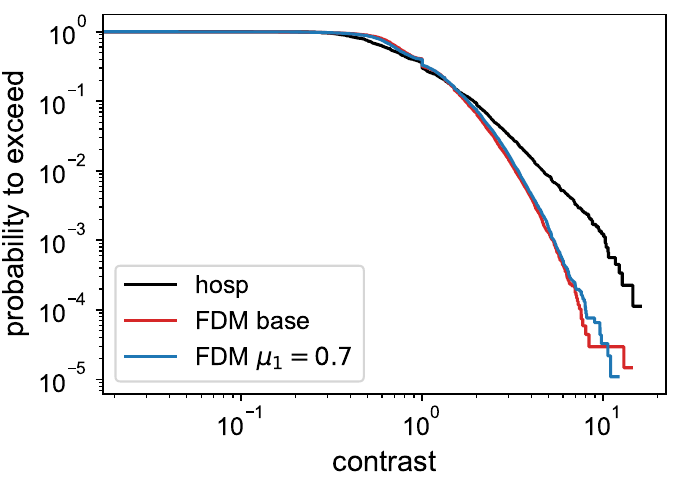}}
\caption{\label{fig:fdm} Distributions (\pte) of (a) the overall contact
  duration, (b) the weights (i.e the sum of the duration of contacts
  for each pair), (c) the number of contacts between each pair, (d)
  their mean interaction time and 
  (e) the duration contrast. The black lines are obtained with the (\hosp)
  data, the red ones with the base FDM implementation and the blue ones
  with the adjusted FDM model $\mu_1=0.7$.
}
\end{figure}

We have also tried varying the $F_0$ and $\mu_2$ parameters but none of the
results were more satisfactory (we did not change $L$ since getting
a correct average degree is an essential feature).

Overall we noticed that the contrast distribution is very robust to
any change of the parameters and can hardly be modified in the default setup.

Each agent in the model is assigned an activity number $r_i$, which gives the
probability to go from an inactive state to an active one. It is sampled
uniformly by default between 0 and 1. We have tried instead to set 
this probability to some fixed values but
again the contrast distribution varied very little.

\newpage
\bibliographystyle{unsrt}
\bibliography{references}